\documentclass[aps,prd,floats,floatfix,preprintnumbers,superscriptaddress,nofootinbib]{revtex4-2}

\usepackage{graphicx} 
\usepackage{epstopdf} 
\usepackage{dcolumn}
\usepackage{amssymb}
\usepackage{amsmath}
\usepackage[hidelinks]{hyperref}
\usepackage{color}
\usepackage{xcolor,colortbl}
\usepackage[utf8]{inputenc}
\usepackage{slashed}

\newcommand{\be}{\begin{equation}}
\newcommand{\ee}{\end{equation}}

\newcommand{\rmd}{\mathrm{d}}
\newcommand{\rme}{\mathrm{e}}
\newcommand{\rmi}{\mathrm{i}}
\newcommand{\calA}{\mathcal{A}}

\newcommand{\calP}{\mathcal{P}}
\newcommand{\calR}{\mathcal{R}}
\newcommand{\calM}{\mathcal{M}}
\newcommand{\wcalM}{\widetilde{\mathcal{M}}}
\newcommand{\calN}{\mathcal{N}}

\newcommand{\calS}{\mathcal{S}}
\newcommand{\calO}{\mathcal{O}}

\newcommand{\calT}{\mathcal{T}}
\newcommand{\calH}{\mathcal{H}}

\newcommand{\qonp}{\boldsymbol{q}_{1\perp}}
\newcommand{\qtwp}{\boldsymbol{q}_{2\perp}}
\newcommand{\qthp}{\boldsymbol{q}_{3\perp}}
\newcommand{\ponp}{\boldsymbol{p}_{1\perp}}
\newcommand{\ptwp}{\boldsymbol{p}_{2\perp}}
\newcommand{\pthp}{\boldsymbol{p}_{3\perp}}
\newcommand{\lonp}{\boldsymbol{l}_{1\perp}}

\newcommand{\rp}{\boldsymbol{r}_{\perp}}

\newcommand{\bp}{\boldsymbol{b}_{\perp}}

\newcommand{\lp}{\boldsymbol{l}_{\perp}}
\newcommand{\ellp}{\boldsymbol{\ell}_{\perp}}
\newcommand{\up}{\boldsymbol{u}_{\perp}}
\newcommand{\vp}{\boldsymbol{v}_{\perp}}
\newcommand{\xp}{\boldsymbol{x}_{\perp}}
\newcommand{\yp}{\boldsymbol{y}_{\perp}}

\newcommand{\qp}{\boldsymbol{q}_{\perp}}

\newcommand{\kp}{\boldsymbol{k}_{\perp}}

\newcommand{\delp}{\boldsymbol{\Delta}_{\perp}}
\newcommand{\dlp}{\boldsymbol{\delta}_{\perp}}

\newcommand{\epsp}{\boldsymbol{\epsilon}_{\perp}}

\newcommand{\pd}{\partial}
\newcommand{\kb}{\boldsymbol{k}}
\newcommand{\sigb}{\boldsymbol{\sigma}}
\newcommand{\Eb}{\boldsymbol{E}}
\newcommand{\lambdab}{\bar{\lambda}}

\newcommand{\Mcc}{{M_0}}
\newcommand{\Dcc}{\Delta_0}
\newcommand{\lth}{\boldsymbol{l}}

\begin{document}
\date{\today}
\preprint{ZTF-EP-24-02}

\title{Photon-Odderon interference in exclusive $\chi_{c}$ charmonium production at the Electron-Ion Collider}
\author{Sanjin Beni\' c}
\affiliation{Department of Physics, Faculty of Science, University of Zagreb, Bijenička c. 32, 10000 Zagreb, Croatia}

\author{Adrian Dumitru}
\affiliation{Department of Natural Sciences, Baruch College, CUNY,
17 Lexington Avenue, New York, NY 10010, USA}
\affiliation{The Graduate School and University Center, The City University of New York, 365 Fifth Avenue, New York, NY 10016, USA}

\author{Abhiram Kaushik}
\affiliation{Centre for Informatics and Computing, Rudjer Bo\v skovi\' c Institute, HR-10002 Zagreb, Croatia}

\author{Leszek Motyka}
\affiliation{Jagiellonian University, Institute of Theoretical Physics,
Łojasiewicza 11, 30-348 Kraków, Poland}

\author{Tomasz Stebel}
\affiliation{Jagiellonian University, Institute of Theoretical Physics,
Łojasiewicza 11, 30-348 Kraków, Poland}

\begin{abstract} 
Exclusive $C=+1$ scalar, axial-vector, and tensor quarkonium production in high-energy electron-proton scattering requires a $C$-odd $t$-channel exchange of a photon or a three gluon ladder. We derive the expressions for the corresponding amplitudes. The relative phase of the photon vs.\ three gluon exchange amplitudes is determined by the sign of the light-front matrix element of the eikonal color current operator $d^{abc}J^{+a}J^{+b}J^{+c}$ at moderate $x$, and is not affected by small-$x$ QCD evolution. Model calculations predict constructive interference, which is particularly strong for momentum transfer $|t|\sim 1$~GeV$^2$ where the cross
section for $\chi_{cJ}$ production exceeds that for pure photon exchange by up to a factor of 4. 
Exclusive $\chi_{cJ}$ electroproduction at the high luminosity Electron-Ion Collider occurs with well measurable rates and measurements of these processes should find evidence for the perturbative Odderon exchange. We also compute the total electroproduction cross section as a function of energy and provide first estimates of the number of $\chi_{cJ}$ events per month at the Electron-Ion Collider design luminosity.
\end{abstract}


\maketitle

\tableofcontents

\section{Introduction}

A basic prediction of QCD, related to the non-zero cubic Casimir invariant, is that there exists a contribution to high-energy
scattering amplitudes from a $C$-conjugation odd exchange driven mostly by gluons, with a weakly energy dependent cross section. 
The possible existence of such an exchange, called the Odderon, was predicted 50 years ago from general principles of quantum field theory~\cite{Lukaszuk:1973nt}.
If the scattering involves hadrons of size much smaller than the QCD confinement scale, or high momentum transfer, then the composition of the exchanged ``state" can be understood from perturbative QCD. At leading order in perturbation theory, it corresponds to an exchange of three gluons which is symmetric under permutations of their colors. At high energies the three gluon exchange is dressed by higher order QCD corrections enhanced by logarithms of energy that have to be resummed into a gluon ladder. It accounts for interactions between the exchanged $t$-channel gluons, and for virtual corrections. The resulting ``reggeized" three gluon exchange is the hard Odderon.  In the weak field limit the corresponding Bartels-Jaroszewicz-Kwieciński-Praszałowicz (BJKP) linear QCD evolution equation has been obtained long ago~\cite{Bartels:1980pe,Kwiecinski:1980wb,Jaroszewicz:1980mq},
and it amounts to the reggeization of the exchanged gluons by iteration of the Balitsky-Fadin-Kuraev-Lipatov (BFKL) interaction kernel~\cite{Lipatov:1976zz,Kuraev:1977fs,Balitsky:1978ic} for each pair of gluons in the ladder. The NLO corrections to the BJKP equation were found in Ref.~\cite{Bartels:2012sw}.
At asymptotically high energies the leading solution to the BJKP equation corresponds to a configuration of gluons in which two reggeized gluons are combined together~\cite{Bartels:1999yt} -- the Bartels–Lipatov–Vacca (BLV) solution. The BLV solution couples to color dipoles so it is particularly important in $C$-even quarkonia (or meson) photo- and electroproduction~\cite{Bartels:1999yt}. The intercept of this solution equals one, in the leading logarithmic approximation~~\cite{Bartels:1999yt}, and it was argued that it remains at this value to all orders in the perturbative expansion~\cite{Stasto:2009bc}.

The non-linear Odderon evolution equation for dipole--proton scattering was established in Refs.~\cite{Kovchegov:2003dm,Hatta:2005as}. This equation typically produces solutions where the Odderon
amplitude decreases with energy~\cite{Motyka:2005ep,Lappi:2016gqe,Yao:2018vcg,Contreras:2020lrh,Benic:2023ybl}, as will be confirmed also in our present work. The seed for this evolution to small $x$ is provided by the cubic Casimir $d^{abc} J^{+a}J^{+b}J^{+c}$ in the effective action describing color current ($J^{+a}$) fluctuations in the proton at moderately small $x$~\cite{Jeon:2005cf,Jeon:2004rk}, which allows the $C$-odd gluon ladder to couple to the proton. The matrix element\footnote{We are here concerned with the off-forward
matrix element for non-zero momentum transfer $t$. In the $t\to0$ limit, instead,
the ``spin dependent Odderon"~\cite{Zhou:2013gsa,Boussarie:2019vmk,Hagiwara:2020mqb} 
is associated with a spin flip of the proton in DIS.}
of this operator could also be evaluated directly from the
light-cone wave function of the proton~\cite{Dumitru:2018vpr,Dumitru:2021tqp}.
A key point for the present analysis is that the Odderon evolution equation
to small $x$ {\em does not alter the sign} of the amplitude. Hence, the
interference pattern with the electromagnetic amplitude due to
single photon exchange is determined by the sign of the
matrix element of the $d^{abc} J^{+a}J^{+b}J^{+c}$ operator at the initial,
moderately small $x$.
\\

The TOTEM collaboration at the CERN-LHC has measured the
differential cross section for $pp$ elastic scattering at $\sqrt{s} =
2.76$~GeV~\cite{TOTEM:2018psk}. They observe a significant
difference to the cross section for $p\bar p$
scattering at $\sqrt{s} = 1.96$~GeV measured
by D0~\cite{D0:2012erd}. Assuming
that the difference in energy is negligible they conclude that these
results provide evidence for the exchange of a color singlet gluonic state
with odd C-parity, the Odderon. However, these 
measurements involve the scattering of hadrons with a size of order
the QCD confinement scale as well as low momentum transfers, $|t| \ll
1$~GeV$^2$. Hence, an analysis of the nature of the $t$-channel
exchange from perturbative QCD would not appear reliable.\\

Exclusive production of pseudo-scalar $\eta_c$ quarkonia in DIS has
previously been proposed as a process suitable for the discovery
of the hard Odderon exchange in QCD~\cite{Schafer:1992pq,Czyzewski:1996bv,Engel:1997cga,Kilian:1997ew,Berger:1999ca,Bartels:2001hw,Dumitru:2019qec,Benic:2023ybl}.
In practice, the detection of this process is difficult as the $\eta_c$ has small branching ratios to all 
relevant decay channels. For example, the radiative $\eta_c \to \gamma \gamma$ mode has a tiny ($\sim 10^{-4}$) branching ratio\footnote{We take all branching ratios and particle masses from the ``Review of Particle Physics"~\cite{ParticleDataGroup:2022pth}.}, while the hadronic $\eta_c \to \rho \rho$ mode has a branching ratio of $1.5\%$.
Also, the cross section for exclusive production of $J/\psi$ is expected to
be far greater than that for $\eta_c$, and so most $\eta_c$'s
would originate from a $J/\psi \to \eta_c \gamma$ decay\footnote{This issue could be mitigated by considering $\eta_c(2S)$ instead, since it lies between the $J/\psi$ and the $\psi(2S)$. The branching ratio $\psi(2S) \to \eta_c + \gamma$ is favourably small ($(7\pm 5)\times 10^{-4}$) but the detection of the $\eta_c(2S)$ is again problematic due to the small branching ratios for many channels.}.
The experimental identification of the soft photon ($M_{J/\psi}=3.097$~GeV,
$M_{\eta_c}=2.984$~GeV) is difficult~\cite{Klein:2018ypk,Harland-Lang:2018ytk}.

Alternative approaches for the discovery of the hard Odderon include Pomeron--Odderon interference (with a background
from Pomeron--photon interference) which would manifest in asymmetries in
exclusive production of two charged particles in DIS~\cite{Brodsky:1999mz,Hagler:2002nh,Hagler:2002nf};
for the $\gamma^* p \to \chi_c p$ process considered here, it is instead
the interference of the Primakoff-like amplitude with the
QCD Odderon exchange amplitude which will be important, i.e.\ photon--Odderon interference.
Another process which involves a $C$-odd $t$-channel
gluonic exchange is exclusive production of a $J/\psi$
in double-diffractive proton-proton scattering~\cite{Schafer:1991na,Bzdak:2007cz}. At high-energy proton colliders this requires instrumentation over a large range of rapidity, close to the beams.

The GlueX Collaboration at Jefferson Lab has recently reported the observation of $56.5\pm 8.2$ exclusive $\chi_{c1}(1P)$ and $12.7\pm 4.5$ $\chi_{c2}(1P)$ production events near threshold energy~\cite{GlueX:2023}. They noticed a ``dramatic difference" in the momentum transfer $t$-distribution of events
as compared to $J/\psi$ production in that the cross section for
$\chi_c$ production appears to drop off much less rapidly with increasing $|t|$.
In other words, the probability that the struck proton remains intact at high momentum transfer
is much greater in $\gamma p \to \chi_c p$ than in $\gamma p \to J/\psi p$ events.
This remarkable result illustrates the importance of the underlying QCD dynamics as opposed to
naive expectations based solely on the mass of the produced quarkonia. A much harder $t$-dependence for exclusive
production of $\eta_c$ vs.\ $J/\psi$ in eikonal dipole--proton scattering
is predicted by simple light-front constituent quark models~\cite{Dumitru:2019qec}.
Inspired by the GlueX measurement of exclusive $\chi_c$ production in DIS at Jefferson Lab, here we consider the same process but at high energies appropriate for the future Electron-Ion Collider 
(EIC)~\cite{Accardi:2012qut,AbdulKhalek:2021gbh} where dipole model factorization \cite{Nikolaev:1990ja,Mueller:1993rr} applies, and 
where the $t$-channel exchange of a $C$-odd color singlet state dominates.

Within the $\chi_{cJ}$ family ($J = 0$, $1$, $2$), $\chi_{c1}$ has the largest branching ratio for the radiative 
$\chi_{c1} \to J/\psi \gamma$ decay ($34.3\%$ as opposed to $1.4\%$ for $\chi_{c0}$ and $19.0\%$ for $\chi_{c2}$). Indeed,
this was the detection channel used by the GlueX measurement. 
Therefore, the production of $\chi_{c1}$ axial-vector and $\chi_{c2}$ tensor quarkonia may prove the most promising channels for discovery
of the hard Odderon. This is further corroborated by the fact that for $\chi_{c1}$ and $\chi_{c2}$ the Odderon and photon exchange contributions
become comparable at lower $|t|$ than for $\chi_{c0}$, within the acceptance of the proton spectrometer of the EIC design detector~\cite{AbdulKhalek:2021gbh}.
However, for the $\chi_{cJ}$ quarkonia there is again a feed-down channel from exclusive production of $\psi(2S)$ with subsequent decay $\psi(2S) \to \chi_{cJ}\gamma$. Hence, the identification and rejection of such feed down is required.

The paper is organized as follows. In Sec.~\ref{sec:ampmain} we start with the computation of the amplitude and the corresponding $\gamma^* p$ cross section for exclusive production of $C$-even scalar, axial vector and tensor quarkonia. We derive the light-cone wave functions of quarkonia, whereas the amplitude is obtained as an overlap with the photon light-cone wave function. We introduce the Odderon exchange amplitude and discuss its evolution to small-$x$. Sec.~\ref{sec:prim} is devoted to the computation of the Primakoff contribution, where we pay special attention to the $t\to 0$ limit. In Sec.~\ref{sec:boosted} we perform a numerical fit of the $\chi_{cJ}$ wave functions. The main results are shown in Sec.~\ref{sec:numer}, where we numerically compute exclusive $\gamma^* p \to \chi_{cJ} p$ and $ep \to \chi_{cJ} ep$ cross sections and the expected number of events at the EIC. Our findings are summarized in the final Sec.~\ref{sec:sum}. Several Appendices follow where we explain the computational steps leading to some of the results from the main text.

\section{The production amplitude of $C$-even quarkonia}
\label{sec:ampmain}

We consider the process $\gamma^*(q) p(P) \to p(P') \calH(\Delta)$ where $\calH$ is a $C$-even quarkonium state, specifically a $P$-wave  $\chi_{cJ}$ charmonium state with $J = 0$, $1$, $2$. However our derivations are
presented in a general form so that
they apply also to other $C = +1$ $q\bar{q}$ bound states such as bottomonia, or even mesons with light flavor content. 

Our computation is performed in the dipole frame where $q = (-Q^2/2q^-,q^-,0,0)$, $P = (P^+,0,0,0)$,
and the proton mass $M_p \ll \sqrt{(q+P)^2}$ is neglected. We use the convention for the components $(v^+,v^-,v^1,v^2)$ of a four vector $v$ where $v^{\pm} = (v^0 \pm v^3)/\sqrt{2}$ are the $\pm$ light-cone components. The amplitude is computed using light-cone gauge $A^-  = A\cdot n = 0$, that is, $n^\mu = \delta^{\mu +}$, for the field corresponding to the incoming photon. 
 
\begin{figure}[htb]
  \begin{center}
  \includegraphics[scale = 0.45]{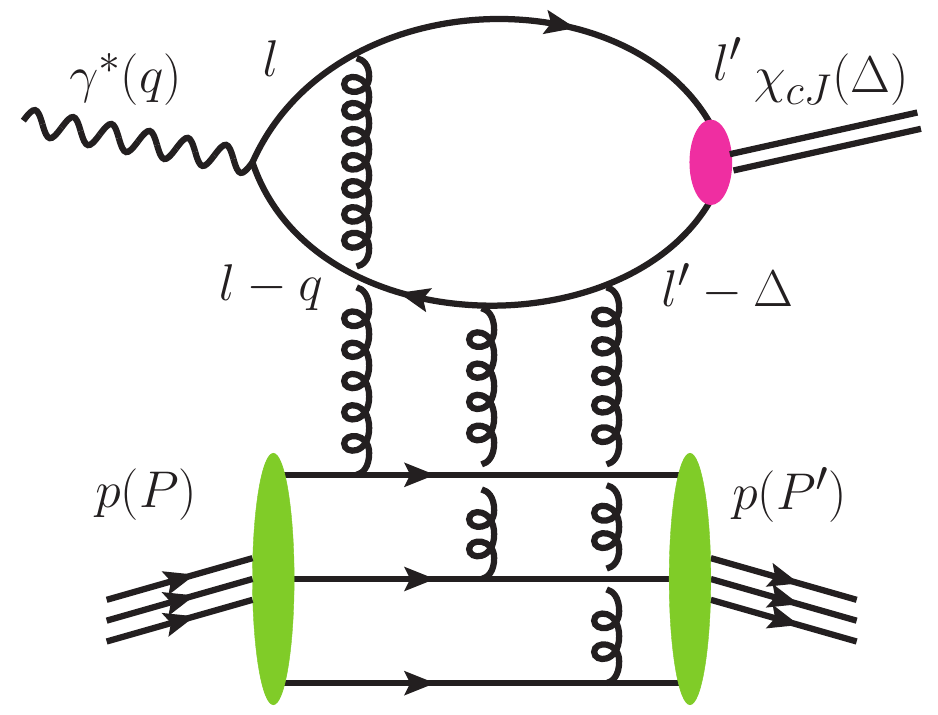}
  \end{center}
  \caption{An example Feynman diagram illustrating the amplitude for exclusive production of $\chi_{cJ}$'s via three gluon (Odderon) exchange from the proton in $\gamma^* p$ collision. The green blobs represent the proton wave function that is shown here in terms of three valence quarks. The magenta blob is the $\chi_{cJ}$ wave function. In the dipole approach the two gluons coupling to the same quark line in the top part of the diagram occupy the same point in transverse position space.
  }
  \label{fig:chicJ}
\end{figure}

Following a similar procedure as in \cite{Dumitru:2019qec,Benic:2023ybl} our starting point is the amplitude 
\be
\left\langle\calM_{\lambda\lambdab}(\gamma^* p \to \calH p)\right\rangle = 2 q^- N_c \int_{\rp\bp} \rme^{-\rmi \delp\cdot \bp} \rmi\calO(\rp,\bp)\calA_{\lambda\lambdab}(\rp,\delp)\,,
\label{eq:M}
\ee
with $\lambda$ and $\lambdab$ the helicities of the incoming photon and the outgoing quarkonium, respectively. For brevity, we will often times write the amplitude simply as $\langle\calM_{\lambda\lambdab}\rangle$ (omitting the round parenthesis). Here $\calA_{\lambda\lambdab}(\rp,\delp)$ is the reduced amplitude, described physically in terms of the photon ($\Psi^{\gamma}_{\lambda,h\bar{h}}$) and the quarkonium ($\Psi^{\calH}_{\lambda,h\bar{h}}$) wave function overlap. We have\footnote{We employ the following abbreviation for the transverse coordinate space integrals: $\int_{\rp} \equiv \int \rmd^2 \rp$. For transverse and longitudinal momentum space integrals we use: $\int_{\lp} \equiv \int \rmd^2 \lp/(2\pi)^2$ and $\int_z \equiv \int_0^1 \rmd z/(4\pi)$.}
\be
\calA_{\lambda\bar{\lambda}}(\rp,\delp) = \int_z \int_{\lp \lp'}\sum_{h\bar{h}} \Psi^{\gamma}_{\lambda,h\bar{h}}(\lp,z)\Psi^{\calH *}_{\bar{\lambda},h\bar{h}}(\lp' - z\delp,z)\rme^{\rmi (\lp - \lp' + \frac{1}{2}\delp)\cdot \rp}\,.
\label{eq:overlapft}
\ee
A particular contribution to the amplitude \eqref{eq:M} is shown in Fig.~\ref{fig:chicJ}, where the upper heavy quark loop represents the reduced amplitude $\calA_{\lambda\bar{\lambda}}(\rp,\delp)$, and the three-gluon contribution is the lowest order depiction of the Odderon exchange amplitude $\calO(\rp,\bp)$ for the particular case where the gluons connect to three different quarks in the proton wave function, represented by the green blob.

The photon wave-function in momentum space is given by
\be
\Psi^{\gamma}_{\lambda,h\bar{h}}(\kp,z) \equiv \sqrt{z\bar{z}}\,
\frac{\bar{u}_h(k) e q_c \slashed{\epsilon}(\lambda,q)v_{\bar{h}}(q-k)}{\kp^2 + \varepsilon^2}\,,
\label{eq:photon}
\ee
where $q_c = +2/3$ is the fractional electric charge of the $c$-quark and $u_h(k)$ ($v_{\bar{h}}(q-k)$) are particle (antiparticle) spinors with $h = \pm 1$ indicating the sign of the helicity $h/2$. For explicit computations we use Lepage-Brodsky (LB) spinors \cite{Lepage:1980fj} - see App.~\ref{sec:photonwf} for their expressions.
$\epsilon(\lambda,q)$ is the photon polarization vector: we have
$\epsilon(0,q) = (Q/q^-,0,0,0)$ for longitudinal ($\lambda = 0$) and 
$\epsilon(\lambda,q) = (0,0,\epsp^\lambda)$ for transverse ($\lambda = \pm 1$) polarization.  We follow the LB convention for the 2D polarization vector: $\epsp^{\lambda} \equiv (-\lambda,-\rmi)/\sqrt{2}$, and the common shorthands $\varepsilon \equiv \sqrt{m_c^2 + z\bar{z}Q^2}$ and $\bar{z} \equiv 1 - z$, with $z = k^- /q^-$. We compute the photon wave function as described in App.~\ref{sec:photonwf}, with the result
\be
\begin{split}
&\Psi^\gamma_{+,h\bar{h}}(\kp,z) = -\sqrt{2} e q_c \left[k_\perp \rme^{\rmi \phi_k} \left(z\delta_{h+}\delta_{\bar{h}-} - \bar{z}\delta_{h-}\delta_{\bar{h}+}\right) + m_c \delta_{h +}\delta_{\bar{h}+}\right]\frac{1}{\kp^2 + \varepsilon^2}\,,\\
&\Psi^\gamma_{-,h\bar{h}}(\kp,z) = -\sqrt{2} e q_c \left[k_\perp \rme^{-\rmi \phi_k} \left(\bar{z}\delta_{h+}\delta_{\bar{h}-} - z\delta_{h-}\delta_{\bar{h}+}\right) + m_c \delta_{h -}\delta_{\bar{h}-}\right]\frac{1}{\kp^2 + \varepsilon^2}\,,\\
&\Psi^\gamma_{0,h\bar{h}}(\kp,z) = \frac{e q_c 2 Q z\bar{z}}{\kp^2 + \varepsilon^2} \delta_{h,-\bar{h}}\,,
\end{split}
\label{eq:photwf}
\ee
where $k_\perp \rme^{\pm\rmi \phi_k} = k^1 \pm \rmi k^2$.
Eq.~\eqref{eq:photwf} coincides with the result of Ref.~\cite{Dosch:1996ss} 
for both $\lambda = 0$ and $\lambda = \pm 1$ after adjusting for 
an overall factor of $-\sqrt{N_c}$ due to the difference in conventions. Comparing to Ref.~\cite{Lappi:2020ufv} the result coincides for 
$\lambda=0$ (again up to a factor $-\sqrt{N_c}$) but there is an opposite overall sign for $\lambda = \pm 1$ (the relative factor being 
$+\sqrt{N_c}$), as also explicitly noted in \cite{Lappi:2020ufv}. While such overall signs play no role for the processes involving 
overlaps of photons and vector quarkonia, the sign is important for tensor quarkonia, since the tensor wave function can be understood as 
a linear superposition of the vector wave functions, see Eq.~\eqref{eq:Elong} below. 
The computation described in App.~\ref{sec:photonwf} takes the LB spinors as a starting point from which we proceed to explicitly compute the photon wave function as well as all the subsequent quarkonia wave functions in a systematic and self-contained fashion.

The explicit expressions for the quarkonia wave functions will be given in the following subsection. Here we want to emphasize that thanks to the $C = +1$ parity of the quarkonium wave function (see Eq.~\eqref{eq:cpar}) below) and to the $C = -1$ parity of the photon, the amplitude is strictly proportional to the $C = -1$ Odderon amplitude $\calO(\rp,\bp)$.
In the high energy limit of eikonal dipole-proton scattering we have \cite{Kovchegov:2003dm,Hatta:2005as}
\be
\calO(\rp,\bp) = -\frac{1}{2\rmi N_c}{\rm tr}\left\langle V(\xp)V^\dag(\yp) - V(\yp)V^\dag(\xp)\right\rangle\,,
\label{eq:odd}
\ee
where $\rp$ is the dipole size and $\bp$ the impact parameter
\be
\rp \equiv \xp - \yp\,,\qquad \bp \equiv \frac{\xp + \yp}{2}\,,
\label{eq:rb}
\ee 
respectively~\footnote{The
impact parameter of the dipole-proton collision is actually $\tilde{\boldsymbol{b}}_\perp = z\xp+\bar z\yp = \bp - (\frac{1}{2}-z)\rp$,
and $\tilde{\boldsymbol{b}}_\perp$ is the Fourier conjugate of the transverse momentum transfer $\delp$.
This is the origin of the ``off-forward phase" $-i\dlp\cdot\rp$ in Eq.~(\ref{eq:redA}).}. 
$V(\xp)$ and $V^\dag(\xp)$ represent Wilson lines at transverse coordinate $\xp$ which describe the propagation
of the quark and anti-quark, respectively, through the $A^+$ color field of the target proton in covariant gauge; 
and $\langle\cdots\rangle$ denotes an average over the configurations of that field, see App.~\ref{sec:Odderon-sign} for the conventions used in this work.

In Fourier space, the Odderon amplitude is given by
\be \label{eq:O(t)}
\calO(\rp,\delp) \equiv \int_{\bp} \rme^{-\rmi \delp \cdot \bp} \calO(\rp,\bp)\,.
\ee
The $C$-invariance of the full amplitude \eqref{eq:M} is now easily verified. We can conventionally start by first exchanging the quark-antiquark coordinates $\xp \leftrightarrow \yp$ so that $\rp \to - \rp$ in the Odderon amplitude. Because of the phases in \eqref{eq:overlapft} this needs to be followed by an exchange of quark and antiquark transverse momenta $\lp \leftrightarrow \lp' - z\delp$ and $z\leftrightarrow \bar{z}$. By this transformation only the photon wave function picks up a sign which cancels with the sign in the transformation of the Odderon amplitude as $\calO(-\rp,\bp) = - \calO(\rp,\bp)$.

\subsection{Light-cone wave function of $C$-even quarkonia}

The light-cone wave function of the $C$-even quarkonia, $\Psi^{\calH}_{\bar{\lambda},h\bar{h}}$, is modeled by the following covariant ansatz
\be
\Psi^{\calH}_{\lambdab,h\bar{h}}(\kp,z) \equiv \frac{1}{\sqrt{z\bar{z}}}\bar{u}_h(k) \Gamma_{\lambdab}^{\calH}(k,k') v_{\bar{h}}(k')\phi_{\calH}(\kp,z)\,.
\label{eq:axmes}
\ee
Here $\lambdab$ is the quarkonium helicity and $k$ ($k' = \Delta_0 - k$) is the quark (antiquark) momentum, with $\Dcc$ the invariant four-momentum of the $c\bar c$ pair. $\Gamma^{\calH}_{\lambdab}$ is the appropriate Dirac matrix vertex function for either a scalar ($\calH = \calS$), axial vector ($\calA$) or a tensor ($\calT$) quarkonium which we take as\footnote{Our convention is $\gamma^5 \equiv +\rmi \gamma^0\gamma^1 \gamma^2 \gamma^3$ and $\epsilon^{0123} \equiv +1 = - \epsilon^{+-12}$.}
\be
\Gamma^{\calH}_{\lambdab}(k,k') = 
\begin{cases}
1~,&  \calH = \calS\,,\\
\rmi \gamma_5 \slashed{E}(\lambdab,\Delta_0)~,& \calH = \calA\,,\\
\frac{1}{2}\left(\gamma_\mu (k_\nu - k'_\nu) + \gamma_\nu (k_\mu - k'_\mu)\right)E^{\mu\nu}(\lambdab,\Delta_0)~, & \calH = \calT\,.
\end{cases}
\label{eq:mesvert}
\ee
$\phi_{\calH}(\kp,z)$ is a non-perturbative scalar function that we model later (see Sec.~\ref{sec:boosted}).

The structure of \eqref{eq:mesvert} is motivated in part to ensure the correct $C$-even property of the quarkonia wave function in \eqref{eq:axmes}. Namely, by exchanging the quark and the antiquark momenta ($k\leftrightarrow k'$) and helicity ($h\leftrightarrow \bar{h}$) we have
\be
\bar{u}_{\bar{h}}(k') \Gamma_{\lambdab}^{\calH}(k',k) v_h(k) = + \bar{u}_h(k) \Gamma^{\calH}_{\lambdab}(k,k') v_{\bar{h}}(k')\,,
\label{eq:cpar}
\ee
which holds thanks to the relation $C (\Gamma^{\calH}_{\lambdab}(k',k))^T C^{-1} = + \Gamma_{\lambdab}^{\calH}(k,k')$ with $C = \rmi \gamma^0 \gamma^2$. Furthermore, the $\Gamma^{\calT}_{\lambdab}(k,k')$ vertex is modelled as a coupling to the fermionic energy-momentum tensor, see e.~g.~\cite{Berger:2000wt,Fillion-Gourdeau:2007adh,Pasechnik:2009qc,Lansberg:2009xh}.

In eq.~\eqref{eq:mesvert} the $E(\lambdab,\Delta_0)$ describe the polarization state of the axial quarkonia. To ensure that the axial vector carries the correct quantum numbers, we require the transversality condition
\be
\Dcc\cdot E(\lambdab,\Dcc) = 0\,.
\label{eq:Ecc}
\ee
This way, in the $c\bar c$ rest frame, $E^\mu(\bar{\lambda},\Delta_0)$ reduces to its spatial components corresponding to the total spin vector of the $c\bar c$ state.  
Recall that in the light-cone formulation the ``$-$'' and $\perp$ components of $\Dcc$ are exactly equal to the corresponding components 
 of the quarkonia four-momentum $\Delta$, but that the ``$+$'' components differ: $\Delta^+ = (M_\calH ^2 + \delp^2)/2\Delta^-$ and  $\Dcc^+ = 
 (\Mcc ^2 + \delp^2)/2\Delta^-$ with $M_0^2 = \Delta_0^2$ being the invariant mass of the $c\bar{c}$ pair and $M^2_\calH = \Delta^2$ the quarkonia mass. This is because the $c\bar c$ pair is allowed to be virtual in the ``$+$'' component, conjugate to the 
 light-cone time. Explicitly, we have
\be
E^\mu(\lambdab = \pm 1,\Dcc) = \left(\frac{\epsp^{\lambdab}\cdot \delp}{\Delta^-},0,\epsp^{\lambdab}\right) \,, \quad
E^\mu(\lambdab = 0,\Dcc) = \left(\frac{\delp^2 - M_0^2}{2 \Mcc \Delta^-},\frac{\Delta^-}{\Mcc},\frac{\delp}{\Mcc}\right) = \frac{1}{\Mcc}\Dcc^\mu - \frac{\Mcc}{\Delta^-}n^\mu\, .
\label{eq:Elong}
\ee
Using $\Delta$ instead of $\Delta_0$ has no effect on the $\lambdab = \pm 1$ components, namely $E^\mu(\lambdab = \pm 1,\Dcc) = E^\mu(\lambdab = \pm 1,\Delta)$. However, the $\lambdab = 0$ case differs as
%
\be
E^\mu(\lambdab = 0,\Delta) = \frac{1}{M_\calA}\Delta^\mu - \frac{M_\calA}{\Delta^-}n^\mu\, .
\label{eq:EAlong}
\ee
We have checked
that the projector $\rmi\gamma_5 \slashed{E}(\lambdab = 0,\Delta)$ (instead of $\rmi\gamma_5 \slashed{E}(\lambdab = 0,\Delta_0)$) leads to incorrect results  for the $c\bar c$ state 
as it leaves an admixture with spin different from 1.
This happens because in the light-cone formulation the rest frames of the quarkonia and of the $c\bar c$ partonic state in general move with different velocities, which is a consequence of the difference in their ``+'' components of the four-momenta. The necessary condition for the correct projector, $\Dcc \cdot E(\lambdab,\Delta) = 0$, is fulfilled if and only if $\Mcc = M_\calA$.
In other words, we will assume that the amplitude for the transition from $E^\mu(\lambdab=0,\Dcc)$ to $E^\mu(\lambdab=0,\Delta)$ is equal to one. A similar approach was used in Ref.~\cite{Cheng:2003sm}. The underlying principle of parton--hadron duality is also the basis for the successful approach to exclusive $\rho^0$ electroproduction~\cite{Martin:1996bp,Martin:1999wb}.

The tensor quarkonia polarizations $E^{\mu\nu}(\lambdab,\Delta_0)$ can be obtained in terms of $E^\mu(\lambdab,\Delta_0)$, using Clebsch-Gordan coefficients. We have \cite{Berger:2000wt}
\be
\begin{split}
& E^{\mu\nu}(\pm 2,\Delta_0) = E^\mu(\pm 1,\Delta_0)E^\nu(\pm 1,\Delta_0)\,,\\
& E^{\mu\nu}(\pm 1,\Delta_0) = \frac{1}{\sqrt{2}}\big(E^\mu(\pm 1,\Delta_0)E^\nu(0,\Delta_0) + E^\mu(0,\Delta_0)E^\nu(\pm 1,\Delta_0)\big)\,,\\
& E^{\mu\nu}(0,\Delta_0) = \frac{1}{\sqrt{6}}\big(E^\mu(+1,\Delta_0)E^\nu(-1,\Delta_0) + E^\mu(-1,\Delta_0)E^\nu(+1,\Delta_0) + 2 E^\mu(0,\Delta_0)E^\nu(0,\Delta_0)\big)\,.\\
\end{split}
\label{eq:tenspol}
\ee
These polarization tensors are traceless, $g_{\mu\nu}E^{\mu\nu}(\lambdab,\Delta_0) = 0$, and symmetric, $E^{\mu\nu}(\lambdab,\Delta_0) = E^{\nu\mu}(\lambdab,\Delta_0)$. We also have transversality conditions, $\Delta_0^\mu E_{\mu\nu}(\lambdab,\Delta_0) = 0 = \Delta_0^\nu E_{\mu\nu}(\lambdab,\Delta_0)$, as a consequence of \eqref{eq:Ecc}. We have checked that Eqs.~\eqref{eq:tenspol} are consistent with the polarization tensors written in \cite{Pasechnik:2009qc}, up to overall signs.

It is instructive to explicitly compute the quarkonia wave functions in this approach. Using LB spinors \cite{Lepage:1980fj} we find for the scalar quarkonia
\be
\Psi^\calS_{h\bar{h}}(\kp,z) = \frac{1}{z\bar{z}}\left[h k_\perp \rme^{- \rmi h\phi_k}\delta_{h\bar{h}} - m_c (z-\bar{z})\delta_{h,-\bar{h}}\right]\phi_\calS(\kp,z)\,.
\label{eq:psiS}
\ee
The helicity structure of the wave function agrees completely with the result in \cite{Babiarz:2020jkh}, see (A.18) there.
For the axial quarkonia we have
\be
\begin{split}
&\Psi^\calA_{+ 1,h\bar{h}}(\kp,z) = -\frac{\sqrt{2}\rmi}{z\bar{z}}\left[k_\perp \rme^{\rmi \phi_k}(\bar{z}\delta_{h+}\delta_{\bar{h}-} +  z \delta_{h-}\delta_{\bar{h}+}) + m_c(z-\bar{z})\delta_{h+}\delta_{\bar{h}+}\right]\phi_{\calA,T}(\kp,z)\,,\\
&\Psi^\calA_{- 1,h\bar{h}}(\kp,z) = -\frac{\sqrt{2}\rmi}{z\bar{z}}\left[k_\perp \rme^{-\rmi \phi_k}(\bar{z}\delta_{h+}\delta_{\bar{h}-} +  z \delta_{h-}\delta_{\bar{h}+}) - m_c(z-\bar{z})\delta_{h-}\delta_{\bar{h}-}\right]\phi_{\calA,T}(\kp,z)\,,\\
&\Psi^\calA_{0,h\bar{h}}(\kp,z) = -\frac{\rmi}{z\bar{z}}\frac{1}{M_\calA}\left(2\kp^2 h\delta_{h,-\bar{h}} + 2m_c k_\perp \rme^{-\rmi h\phi_k}\delta_{h\bar{h}}\right)\phi_{\calA,L}(\kp,z)\,,
\end{split}
\label{eq:axt0}
\ee
where $k_\perp \rme^{\pm \rmi \phi_k} \equiv k^1 \pm \rmi k^2$. {At this point we have introduced different scalar functions $\phi_{\calA,T}(r_\perp,z)$ and $\phi_{\calA,L}(r_\perp,z)$ for transversely and longitudinally polarized quarkonia\footnote{For the sake of simplicity, in the case of $\lambdab=0$, we have redefined $\phi_{\calA,L}(\kp,z) / M_0 \to \phi_{\calA,L}(\kp,z) / M_\calA$, which introduces only subleading effects in the heavy quark limit.}.}
The helicity structure of \eqref{eq:axt0} coincides with the axial part of the $Z$-boson wave function from Refs.~\cite{Fiore:2005yi,Motyka:2008ac}. In Ref.~\cite{Babiarz:2022xxm} the axial quarkonia wave function was computed starting from the quarkonia rest frame followed by a Melosh transformation. The resulting wave functions, their Eqs.~(A.7) and (A.9), agree up to an overall normalization constant with \eqref{eq:axt0}.

For the tensor quarkonia we obtain the following results
\be
\begin{split}
&\Psi^\calT_{+2,h\bar{h}}(\kp,z) = -\frac{2}{z\bar{z}} k_\perp \rme^{\rmi \phi_k}\left[k_\perp \rme^{\rmi \phi_k}(z\delta_{h +}\delta_{\bar{h}-} - \bar{z}\delta_{h-}\delta_{\bar{h} +}) + m_c \delta_{h+}\delta_{\bar{h}+}\right]\phi_{\calT,T2}(\kp,z)\,,\\
&\Psi^\calT_{-2,h\bar{h}}(\kp,z) = \frac{2}{z\bar{z}} k_\perp \rme^{-\rmi \phi_k}\left[k_\perp \rme^{-\rmi \phi_k}(\bar{z}\delta_{h +}\delta_{\bar{h}-} - z\delta_{h-}\delta_{\bar{h} +}) + m_c \delta_{h-}\delta_{\bar{h}-}\right]\phi_{\calT,T2}(\kp,z)\,,\\
&\Psi^\calT_{+1,h\bar{h}}(\kp,z) = \frac{M_\calT}{z\bar{z}}\left[-k_\perp \rme^{\rmi\phi_k}\left((3z-4z^2)\delta_{h+}\delta_{\bar{h}-} + (3\bar{z}-4\bar{z}^2)\delta_{h-}\delta_{\bar{h}+}\right) + m_c (z-\bar{z})\delta_{h+}\delta_{\bar{h}+}\right]\phi_{\calT,T}(\kp,z)\,,\\
&\Psi^\calT_{-1,h\bar{h}}(\kp,z) = \frac{M_\calT}{z\bar{z}}\left[k_\perp \rme^{-\rmi\phi_k}\left((3\bar{z}-4\bar{z}^2)\delta_{h+}\delta_{\bar{h}-} + (3z - 4z^2)\delta_{h-}\delta_{\bar{h}+}\right) + m_c (z-\bar{z})\delta_{h-}\delta_{\bar{h}-}\right]\phi_{\calT,T}(\kp,z)\,,\\
&\Psi^\calT_{0,h\bar{h}}(\kp,z) = \frac{\sqrt{2}}{\sqrt{3}}\frac{1}{z\bar{z}}\left[(3 \kp^2 + 2m_c^2)(z-\bar{z})\delta_{h,-\bar{h}} + m_c(k_\perp \rme^{-\rmi \phi_k}\delta_{h+}\delta_{\bar{h}+} - k_\perp \rme^{\rmi \phi_k}\delta_{h-}\delta_{\bar{h}-})\right]\phi_{\calT,L}(\kp,z)\,.\\
\end{split}
\label{eq:tenswf}
\ee
The three different scalar functions $\phi_{\calT,T2}(\kp,z)$, $\phi_{\calT,T}(\kp,z)$ and $\phi_{\calT,L}(\kp,z)$ account
for the two possible transverse and longitudinal polarizations of the tensor quarkonia\footnote{In the case $\lambdab = \pm 1$ we have redefined $M_0 \phi_{\calT,T}(\kp,z) \to  M_\calT \phi_{\calT,T}(\kp,z)$.}. Comparing to eq.~(13) in Ref.~\cite{Berger:2000wt}, the wave function for $\lambdab = + 2$ polarization agrees with the first line in \eqref{eq:tenswf} up to an overall normalization constant. For $\lambdab = -2$ we find an opposite relative sign between the two terms in
square brackets (we agree, though, with the relative sign for $\lambdab = -2$ in eq.~(B.3) of \cite{Berger:2000wt}). For $\lambdab = \pm 1$ and $\lambdab = 0$ the results agree up to an overall normalization constant.

At this point we comment on the spin-momentum structure of the wave functions in the rest frame. 
The $C$ parity of a $q\bar{q}$ bound state is $C = (-1)^{L+S}$ and the $\chi_{cJ}$ quarkonia with $J = 0$, $1$ and $2$
have been classified by the Particle Data Group as
$P$-waves\footnote{For the $J=2$ tensor meson an
$F$-wave component is possible, in principle. Due to the
small velocities of the quarks in the rest frame of the
meson, this component should be small, to agree with
the PDG classification of the $\chi_{c2}$ as a $P$-wave.} ($L = 1$)~\cite{ParticleDataGroup:2022pth}. Also, they are spin triplets, $S = 1$.
Based on our model wave function \eqref{eq:axmes} we therefore explicitly checked that, after replacing the light-cone LB 
spinors with Dirac spinors ($u_h(k) \to u_s(k)$ and $v_h(k) \to v_s(k)$), the wave functions in the rest frame take their 
expected non-relativistic ${}^{3} P_J$ spin-momentum structure. 
The scalar wave function is proportional to $\xi^\dag_{\bar{s}} (\sigb\cdot\kb) \tilde{\xi}_{s}$, 
while the axial-vector meson is $\xi^\dag_{\bar{s}} \sigb\cdot (\kb\times\Eb)\tilde{\xi}_s$ 
\cite{Ji:1992yf,Cheng:2003sm,Babiarz:2020jkh,Babiarz:2022xxm}. Here $\xi_s$ are the standard two-component Pauli spinors and $\tilde{\xi}_s = \rmi \sigma_2 \xi_s^*$. 
For the tensor wave function we obtain $\xi^\dag_{\bar{s}} E^{ij}\sigma_i k_j  \tilde{\xi}_s$ \cite{Gupta:1996ak},
at leading order in the non-relativistic limit.

\subsection{Final expressions for the amplitudes}
\label{sec:FinalAmplitudes}

For explicit computations it is convenient to write the reduced amplitude \eqref{eq:overlapft} in the following equivalent form, see 
e.g.~Refs.~\cite{Dosch:1996ss,Mantysaari:2020lhf,Benic:2023ybl}
\be
\calA_{\lambda\lambdab}(\rp,\delp) = e q_c\int_z \rme^{-\rmi \dlp\cdot\rp}\int_{\lp}\frac{\rme^{\rmi \lp\cdot\rp}}{\lp^2 + \varepsilon^2}\int_{\lp'}\rme^{-\rmi (\lp' - z\delp)\cdot \rp}\phi_\calH(\lp' - z\delp,z)\frac{1}{z\bar{z}}A_{\lambda\lambdab}(\lp,\lp' - z \delp,z)\,,
\label{eq:redA}
\ee
where $\dlp = \frac{1}{2}(z-\bar z)\delp = (z-\frac{1}{2})\delp$. The helicity sum in \eqref{eq:overlapft} is turned into a covariant Dirac trace
\be
A_{\lambda\bar{\lambda}}(\lp,\lp' - z \delp,z) = \frac{1}{(2 q^-)^2}{\rm tr}\left[(\slashed{l} + m_c)\slashed{\epsilon}(\lambda,q)(\slashed{l}-\slashed{q} + m_c)\gamma^- (\slashed{l}' - \slashed{\Delta} + m_c)\Gamma_{\lambdab}^{\calH *}(l',\Delta-l')(\slashed{l}' + m_c)\gamma^-\right]\,,
\label{eq:dirtr}
\ee
containing the physical information on the polarization dependent part of the photon-quarkonia wave function overlap. Together with Eq.~\eqref{eq:M}, Eqs.~\eqref{eq:redA} and \eqref{eq:dirtr} comprise the main formulas that will be used below to write
the  $\gamma^* p \to \calH p$ amplitudes in a form suitable for numerical computations. For the explicit evaluation of the traces we have used FeynCalc \cite{Shtabovenko:2020gxv}.


We start with the computation of the amplitude for scalar quarkonium. Inserting the first line of \eqref{eq:mesvert} into \eqref{eq:dirtr} yields
\be
\begin{split}
& A_{\lambda = 0}(\lp,\lonp,z) = -4 m_c Qz\bar{z}(z-\bar{z})\,,\\
& A_{\lambda = \pm 1}(\lp,\lonp,z) = -2 m_c \left[(z-\bar{z})^2 (\epsp^\lambda\cdot \lp) - (\epsp^\lambda\cdot \lonp)\right]\,,\\
\end{split}
\label{eq:trscal}
\ee
where $\lonp \equiv \lp' - z\delp$. As a cross check, note that the traces \eqref{eq:trscal} are odd under joint $\lp \to - \lp$, $\lonp \to -\lonp$ and $z\to \bar{z}$ transformation. In other words, the reduced amplitude $\calA_\lambda(\rp,\delp)$ in \eqref{eq:redA} is odd under $\rp \to - \rp$. This is consistent with
$\calO(\rp,\delp)$ being odd under $\rp \to - \rp$ in the full amplitude \eqref{eq:M}.

We now plug \eqref{eq:trscal} into \eqref{eq:redA} and Fourier transform to coordinate space. After separating the explicit polarization dependence, the reduced amplitudes are found to be 
\be
\begin{split}
&\calA_0(\rp,\delp) = e q_c \int_z \rme^{-\rmi \dlp\cdot \rp} \calA_L(r_\perp)\,,\\
&\calA_{\lambda = \pm 1}(\rp,\delp) = e q_c \lambda \rme^{\rmi \lambda\phi_r}\int_z \rme^{-\rmi \dlp\cdot \rp}\calA_T(r_\perp)\,,\\
\end{split}  
\label{eq:reducescal}
\ee
where
\be
\begin{split}
&\calA_L(r_\perp) \equiv -\frac{2}{\pi} m_c Q (z-\bar{z}) K_0(\varepsilon r_\perp)\phi_\calS(r_\perp,z)\,,\\
&\calA_T(r_\perp) \equiv \frac{\rmi\sqrt{2}}{2\pi}\frac{m_c}{z\bar{z}}\left[(z-\bar{z})^2\varepsilon K_1(\varepsilon r_\perp)\phi_\calS(r_\perp,z) - K_0(\varepsilon r_\perp) \frac{\pd\phi_\calS}{\pd r_\perp}\right]\,.
\end{split}
\label{eq:Ampscal}
\ee
In the amplitude \eqref{eq:M} we are left with the $\rp$ integral and the $z$ integral. We use the Jacobi-Anger expansion to compute the $\phi_r$ integral and find
\be
\begin{split}
\left\langle \calM_{\lambda = 0}\right\rangle &\equiv q^- \wcalM_{\lambda = 0} \equiv q^- \wcalM_L\,,\\
\left\langle \calM_{\lambda = \pm 1}\right\rangle &\equiv q^- \wcalM_{\lambda = \pm 1} \equiv q^- \lambda \rme^{\rmi \lambda \phi_\Delta} \wcalM_T\,,\\
\end{split}
\label{eq:Mscal}
\ee
where in the first equality we have separated the flux factor and in the second equality the explicit polarization dependence (together with the overall phase). The remainder consists of two different scalar functions
\be
\begin{split}
&\wcalM_L = 8\pi N_c e q_c\sum_{k=0}^\infty (-1)^k \int_z \int_0^\infty r_\perp \rmd r_\perp \calO_{2k+1}(r_\perp,\Delta_\perp) \calA_L(r_\perp)\, {\rm sgn}(z-\bar{z})J_{2k+1}(r_\perp\delta_\perp)\,,\\
&\wcalM_T = 4\pi\rmi N_c e q_c\sum_{k=0}^\infty (-1)^k \int_z \int_0^\infty r_\perp \rmd r_\perp \calO_{2k+1}(r_\perp,\Delta_\perp) \calA_T(r_\perp)\left[J_{2k}(r_\perp\delta_\perp) -J_{2k+2}(r_\perp\delta_\perp)\right]\,,
\end{split}
\label{eq:Mpolscal}
\ee
which are our main expressions to be used in the numerical computations in Sec.~\ref{sec:numer}. The sign function 
appears due to the definition of $\delta_\perp$ as the modulus of the vector $\dlp$: $\delta_\perp = \frac{1}{2}|z-\bar z| \Delta_\perp$.

In the limit $\Delta_\perp \to 0$ the above amplitudes scale as
\be \label{eq:S-forward_amplitude}
\begin{split}
&\wcalM_L\Big|_{\Delta_\perp \to 0} \sim \Delta_\perp^2~,~~~~
\wcalM_T \Big|_{\Delta_\perp \to 0} \sim \Delta_\perp~.
\end{split}
\ee
The scaling exponents are universal, they do not depend on specific details of the Odderon
exchange amplitude $\calO(\rp,\bp)$.

The quantities $\calO_{2k+1}(r_\perp,\Delta_\perp)$ appearing in Eq.~\eqref{eq:Mpolscal} are the azimuthal harmonics of the Odderon amplitude,
\be
\calO_{2k+1}(r_\perp,\Delta_\perp) = 2\pi \rmi (-1)^{k+1} \int_0^\infty b_\perp \rmd b_\perp \calO_{2k+1}(r_\perp,b_\perp)J_{2k+1}(\Delta_\perp b_\perp)\,,
\label{eq:oddft}
\ee
where $\calO_{2k+1}(r_\perp,b_\perp)$ is extracted from $\calO(\rp,\bp)$ as its Fourier series coefficients
\be
\calO(\rp,\bp) = 2\sum_{k = 0}^\infty \calO_{2k+1}(r_\perp,b_\perp)\cos((2k+1)\phi_{rb})\,,\qquad \calO_{2k+1}(r_\perp,b_\perp) = \frac{1}{2\pi}\int_0^{2\pi} \rmd \phi_{rb} \calO(\rp,\bp) \cos((2k+1)\phi_{rb})\,,
\label{eq:omom}
\ee
with $\phi_{rb} = \phi_r - \phi_b$.\\


The calculation for axial vector quarkonia follows the same steps as for scalar case. Detailed derivations of the formulas can be found in App.~\ref{appendix_overlaps_axial_tensor}. We end up with three different scalar functions:
\be
\begin{split}
&\wcalM_B = 4\pi\rmi N_c e q_c\sum_{k=0}^\infty (-1)^k \int_z \int_0^\infty r_\perp \rmd r_\perp \calO_{2k+1}(r_\perp,\Delta_\perp) \calA_B(r_\perp)\left[J_{2k}(r_\perp\delta_\perp) - J_{2k+2}(r_\perp\delta_\perp)\right]\,, \qquad B = TL, LT\\
&\wcalM_{TT} = 8\pi N_c e q_c\sum_{k=0}^\infty (-1)^k \int_z \int_0^\infty r_\perp \rmd r_\perp \calO_{2k+1}(r_\perp,\Delta_\perp) \calA_{TT}(r_\perp)\, {\rm sgn}(z-\bar{z})J_{2k+1}(r_\perp\delta_\perp)\,,
\end{split}
\label{eq:Mpol}
\ee
with respective reduced amplitudes
\be
\begin{split}
&\calA_{LT}(r_\perp) \equiv \frac{\sqrt{2}}{\pi}Q K_0(\varepsilon r_\perp)\frac{\pd\phi_{\calA,T}}{\pd r_\perp}\,,\\
&\calA_{TL}(r_\perp) \equiv \frac{\sqrt{2}}{\pi}\frac{1}{z\bar{z}}\frac{1}{M_\calA}\left[- m_c^2 K_0(\varepsilon r_\perp)\frac{\pd\phi_{\calA,L}}{\pd r_\perp} + \varepsilon K_1(\varepsilon r_\perp)\boldsymbol{\nabla}^2_\perp \phi_{\calA,L}
\right]\,,\\
&\calA_{TT}(r_\perp) \equiv  -\frac{\rmi}{\pi}\frac{z-\bar{z}}{z\bar{z}}\left[\frac{\partial \phi_{\calA,T}}{\partial r_\perp} \varepsilon K_1(\varepsilon r_\perp) - m_c^2 K_0(\varepsilon r_\perp)\phi_{\calA,T}\right]\,.
\end{split}
\label{eq:Amps}
\ee
Here the notation $\wcalM_{TT}$ stands for the amplitude with transition from transversely polarized photon to transversely polarized axial quarkonia. In principle, both the polarization preserving and polarization flipping transition would be possible. However, we find that the polarization flipping transition vanishes, and so $\wcalM_{TT}$ describes only a polarization preserving transition. Similar notation is used for other contributing amplitudes. 
When both, the photon and the axial-vector quarkonium are longitudinally polarized, the corresponding
amplitude also vanishes, see App.~\ref{appendix_overlaps_axial_tensor} for more details.

In the near forward limit, the axial-vector amplitudes scale as
\be \label{eq:A-forward_amplitude}
\wcalM_{TL,LT} \Big|_{\Delta_\perp\to0} \sim \Delta_\perp~,~~~
\wcalM_{TT} \Big|_{\Delta_\perp\to0} \sim \Delta_\perp^2~.
\ee
\\

The procedure is again similar for tensor quarkonia, with detailed steps to be found in App.~\ref{appendix_overlaps_axial_tensor}. The final results for the amplitudes are
\be
\begin{split}
\wcalM_B = -4\pi N_c e q_c\sum_{k=0}^\infty (-1)^k & \int_z \int_0^\infty r_\perp \rmd r_\perp \calO_{2k+1}(r_\perp,\Delta_\perp) \calA_B(r_\perp)\\
& \times {\rm sgn}(z-\bar{z})\left[J_{2k+3}(r_\perp\delta_\perp) + J_{2k-1}(r_\perp\delta_\perp)\right]\,,\quad B = LT2,TTf\,,\\
\wcalM_B = 4\pi \rmi N_c e q_c\sum_{k=0}^\infty (-1)^k & \int_z \int_0^\infty r_\perp \rmd r_\perp \calO_{2k+1}(r_\perp,\Delta_\perp)\calA_B(r_\perp)\\
& \times\left[J_{2k}(r_\perp\delta_\perp) - J_{2k+2}(r_\perp\delta_\perp)\right]\,,\quad B = TT2p,LT,TL\,,\\
\wcalM_B = 8\pi N_c e q_c\sum_{k=0}^\infty (-1)^k & \int_z \int_0^\infty r_\perp \rmd r_\perp \calO_{2k+1}(r_\perp,\Delta_\perp)\calA_B(r_\perp){\rm sgn}(z-\bar{z})J_{2k+1}(r_\perp\delta_\perp)\,,\quad B = TTp,LL\,,\\
\wcalM_{TT2f} = 4\pi \rmi N_c e q_c\sum_{k=0}^\infty (-1)^k & \int_z \int_0^\infty r_\perp \rmd r_\perp \calO_{2k+1}(r_\perp,\Delta_\perp)\calA_{TT2f}(r_\perp)\left[J_{2k+4}(r_\perp\delta_\perp) - J_{2k-2}(r_\perp\delta_\perp)\right]\,,\\
\end{split}
\label{eq:MpolTens}
\ee
where
\be
\begin{split}
&\calA_{LT2}(r_\perp) \equiv \frac{2}{\pi}(z-\bar{z}) Q K_0(\varepsilon r_\perp)\left(\frac{\pd^2 \phi_{\calT,T2}}{\pd r_\perp^2} - \frac{1}{r_\perp} \frac{\pd \phi_{\calT, T2}}{\pd r_\perp}\right)\,,\\
&\calA_{TT2p}(r_\perp) \equiv -\frac{\rmi \sqrt{2}}{\pi}\frac{1}{z\bar{z}}\left((z^2 + \bar{z}^2)\varepsilon K_1(\varepsilon r_\perp)\left(\frac{\pd^2 \phi_{\calT,T2}}{\pd r_\perp^2} - \frac{1}{r_\perp} \frac{\pd \phi_{\calT, T2}}{\pd r_\perp}\right) + m_c^2 K_0(\varepsilon r_\perp) \frac{\pd  \phi_{\calT,T2}}{\pd r_\perp}\right)\,,\\
&\calA_{TT2f}(r_\perp) \equiv \frac{\rmi 2\sqrt{2}}{\pi} \varepsilon K_1(\varepsilon r_\perp)\left(\frac{\pd^2 \phi_{\calT,T2}}{\pd r_\perp^2} - \frac{1}{r_\perp} \frac{\pd \phi_{\calT, T2}}{\pd r_\perp}\right)\,,\\
&\calA_{LT}(r_\perp) \equiv -\frac{\rmi}{\pi}Q M_\calT \left(3-4(z^2 + \bar{z}^2)\right)K_0(\varepsilon r_\perp)\frac{\pd \phi_{\calT,T}}{\pd r_\perp}\,,\\
&\calA_{TTp}(r_\perp) = - \frac{M_\calT}{\sqrt{2}\pi}\frac{z-\bar{z}}{z\bar{z}}
\left[m_c^2K_0(\varepsilon r_\perp)\phi_{\calT,T}(r_\perp,z)  - (z - \bar{z})^2\varepsilon K_1(\varepsilon r_\perp)\frac{\pd \phi_{\calT,T}}{\pd r_\perp}\right]\,,\\
&\calA_{TTf}(r_\perp) = -\frac{2\sqrt{2} M_\calT}{\pi} (z-\bar{z}) \varepsilon K_1(\varepsilon r_\perp)\frac{\pd \phi_{\calT,T}}{\pd r_\perp}\,,\\
&\calA_{LL}(r_\perp) =
-\frac{2\sqrt{2}Q}{\sqrt{3}\pi}(z-\bar{z})K_0(\varepsilon r_\perp)\left(3\boldsymbol{\nabla}_\perp^2 
- 2 m_c^2\right) \phi_{\calT,L}(r_\perp,z)\,,\\
&\calA_{TL}(r_\perp) = \frac{\rmi}{\pi\sqrt{3}} \frac{1}{z\bar{z}}\left[\varepsilon K_1(\varepsilon r_\perp)(z-\bar{z})^2
\left(3\boldsymbol{\nabla}_\perp^2 - 2 m_c^2\right) \phi_{\calT,L}
- m_c^2 K_0(\varepsilon r_\perp)\frac{\pd \phi_{\calT,L}}{\pd r_\perp}\right]\,.
\end{split}
\label{eq:redcT2}
\ee
In this case the $LL$-type transition is allowed. Also, both the polarization preserving and the polarization flipped contributions are allowed, thus explaining the above used notation: $p$ = preserving, $f$ = flipped.

In the near forward limit, the tensor meson amplitudes scale as
\be \label{eq:T-forward_amplitude}
\wcalM_{LT2,TTf} \Big|_{\Delta_\perp\to0} \sim \Delta_\perp^2~,~~~
\wcalM_{TT2p,LT,TL} \Big|_{\Delta_\perp\to0} \sim \Delta_\perp~,~~~
\wcalM_{TTp,LL} \Big|_{\Delta_\perp\to0} \sim \Delta_\perp^2~,~~~
\wcalM_{TT2f} \Big|_{\Delta_\perp\to0} \sim \Delta_\perp^3~.
\ee
These amplitudes, as well as those for scalar and axial vector mesons given in eqs.~(\ref{eq:S-forward_amplitude},\ref{eq:A-forward_amplitude}), vanish for $\Delta_\perp\to0$ at least as fast as\footnote{We thank the
anonymous referee for drawing our attention to this point.}
\be
\calM_{\lambda\bar\lambda} \Big|_{\Delta_\perp\to0} \sim \Delta_\perp^{|\lambda-\bar\lambda|}~.
\ee
Such rule emerges when
the helicity change in the $t$-channel at $t\to0$ is due to the $z$-component of the angular momentum of the $t$-channel state. The rule then follows from the conservation of the angular momentum along $z$. However,
the conservation of the $z$-component of the angular momentum provides only a lower limit on the power of
$\Delta_\perp$. The amplitude cannot decrease any slower in the forward limit but it can decrease faster, i.e.\ when the coefficient of the relevant tensor obtained from an expansion of the amplitudes in $\Delta^i$ vanishes independently of the "canonical" scaling of this tensor.

\subsection{The $\gamma^*p\to \calH p$ cross section}

Using the $\wcalM_{\lambda\lambdab}$ amplitudes obtained in the previous Sec.~\ref{sec:FinalAmplitudes} for the process $\gamma^*(q) p(P) \to \calH(\Delta) p(P')$, the cross section for longitudinal ($L$) and transverse ($T$) photons is given by
\be
\begin{split}
& \frac{\rmd \sigma_L (\gamma^* p \to \calH p)}{\rmd |t|} = \frac{1}{16\pi}\sum_{\lambdab = -J}^J\left| \wcalM_{\lambda = 0,\bar{\lambda}}(\gamma^* p \to \calH p)\right|^2\,,\\
& \frac{\rmd \sigma_T (\gamma^* p \to \calH p)}{\rmd |t|} = \frac{1}{16\pi}\frac{1}{2}\sum_{\lambda = \pm 1}\sum_{\lambdab = -J}^J \left| \wcalM_{\lambda\bar{\lambda}}(\gamma^* p \to \calH p)\right|^2\,.
\end{split}
\label{eq:csmain}
\ee
Inserting the scalar quarkonium amplitudes \eqref{eq:Mscal} into \eqref{eq:csmain} gives
\be
\begin{split}
&\frac{\rmd \sigma_L (\gamma^* p \to \calS p)}{\rmd |t|} = \frac{1}{16\pi}\left| \wcalM_L\right|^2\,,\\
&\frac{\rmd \sigma_T (\gamma^* p \to \calS p)}{\rmd |t|} = \frac{1}{16\pi}\left|\wcalM_T\right|^2\,,
\label{eq:csS}
\end{split}
\ee
where $\wcalM_B$ ($B = L$, $T$) are given in Eqs.~(\ref{eq:Mpolscal}). For axial vector quarkonia we find 
\be
\begin{split}
&\frac{\rmd \sigma_L(\gamma^* p \to \calA p)}{\rmd |t|} = \frac{1}{8\pi}\left| \wcalM_{LT}\right|^2\,,\\
&\frac{\rmd \sigma_T(\gamma^* p \to \calA p)}{\rmd |t|} = \frac{1}{16\pi}\left(\left|\wcalM_{TL}\right|^2 + \left| \wcalM_{TT}\right|^2\right)\,,
\end{split}
\label{eq:csA}
\ee
with $\wcalM_B$ ($B = LT$, $TL$, $TT$) given in \eqref{eq:Mpol}. We have used \eqref{eq:Ms} (see Appendix \ref{appendix_overlaps_axial_tensor}) to relate $\wcalM_{\lambda\lambdab}$ to $\wcalM_A$. For the tensor quarkonia, the cross sections are
\be
\begin{split}
&\frac{\rmd \sigma_L(\gamma^* p \to \calT p)}{\rmd |t|} = \frac{1}{16\pi}\left(\left| \wcalM_{LL}\right|^2 + 2\left| \wcalM_{LT}\right|^2 + 2\left| \wcalM_{LT2}\right|^2\right)\,,\\
&\frac{\rmd \sigma_T(\gamma^* p \to \calT p)}{\rmd |t|} = \frac{1}{16\pi}\left(\left|\wcalM_{TL}\right|^2 + \left| \wcalM_{TTp}\right|^2 + \left| \wcalM_{TTf}\right|^2 + \left| \wcalM_{TT2p}\right|^2 + \left| \wcalM_{TT2f}\right|^2\right)\,,
\end{split}
\label{eq:csT}
\ee
with $\wcalM_B$ ($B$ = $LL$, $LT$, $LT2$, $TL$, $TTp$, $TTf$, $TT2p$, $TT2f$) collected in \eqref{eq:MpolTens} and the relationship to $\wcalM_{\lambda\lambdab}$ in \eqref{eq:AmpsTens}.

\subsection{Odderon amplitude and its evolution with $x$}  \label{sec:Odderon-Evol}

Here we discuss the Odderon amplitude $\calO(\rp,\bp)$ used later in Sec.~\ref{sec:numer} for numerical predictions of
the cross section for exclusive $\chi_{c J}$ production. The initial $\calO(\rp,\bp)$ at $x_\calP=x_0=0.01$ is approximated by the matrix element of the $C$-odd three gluon operator.
This matrix element has been evaluated numerically and tabulated in Ref.~\cite{Dumitru:2022ooz} using
a phenomenological non-perturbative three quark light-cone wave function~\cite{Schlumpf:1992vq,Brodsky:1994fz} 
supplemented by the first correction of perturbative QCD~\cite{Dumitru:2021tqp}. We refer to these references for further details.
Specifically, we
use the table for the first azimuthal harmonic $a_1(r_\perp,b_\perp)$
of the $\calO(\rp,\bp)$ defined in Eq.~(5) of Ref.~\cite{Dumitru:2022ooz}. As explained in 
the App.~\ref{sec:Odderon-sign} the definition of $\calO(\rp,\bp)$ in \cite{Dumitru:2022ooz} is the same as in this work\footnote{However, 
our own numerical investigation and private communication with the authors of Ref.~\cite{Dumitru:2022ooz} revealed that their
plots of $\calO(\rp,\bp)$ and their tabulated values actually correspond to minus the function defined in their Eq.~(3) and in our
Eq.~(\ref{eq:Orb_G3}). We therefore reversed the signs of their table entries for $a_1(r_\perp,b_\perp)$.}. 
The first azimuthal moment is defined in \cite{Dumitru:2022ooz} through $\calO(\rp,\bp) = \cos(\phi_{rb}) a_1(r_\perp,b_\perp) + \dots$. 
Thus, from \eqref{eq:omom} we have the relationship
\be
\calO_1(r_\perp,b_\perp) = \frac{1}{2}a_1(r_\perp,b_\perp)\,.
\ee
We do not account for higher order harmonics such as $\calO_3(r_\perp,b_\perp)$ which has been determined to be
very small~\cite{Dumitru:2022ooz} at $x_0$ and suppressed further by evolution \cite{Lappi:2016gqe,Benic:2023ybl}. Presently, yet higher Fourier components are unknown; we expect
their effects to be well within the current uncertainty of the leading Fourier harmonic.

To determine $\calO(\rp,\bp)$ for $x_\calP<x_0$ we solve the impact parameter dependent~\cite{Golec-Biernat:2003naj,Berger:2010sh} extension of the Balitsky-Kovchegov (BK) equation  (with running coupling kernel -- rcBK) to the coupled non-linear evolution equations for the Pomeron and the Odderon derived in Refs.~\cite{Kovchegov:2003dm,Hatta:2005as} and analyzed in Refs.~\cite{Motyka:2005ep,Lappi:2016gqe,Yao:2018vcg}. These equations describe the evolution
with rapidity $Y = \ln(x_0/x_{\calP})$
of the real and imaginary parts of the dipole $S$-matrix. 
For the real part (the Pomeron), we use the initial condition from \cite{Lappi:2013zma}. The coupled BK equations for the Pomeron-Odderon system are solved in the ``local approximation"~\cite{Kowalski:2008sa,Lappi:2013zma},
where the impact parameter $\bp$ becomes an external parameter. This approximation may be
less justified for the Odderon than for the Pomeron since the former amplitude peaks at smaller $b_\perp$~\cite{Dumitru:2022ooz}.
The $t$-dependence of three gluon exchange is then obtained via
Fourier transform, Eq.~(\ref{eq:O(t)}). Technical details of the implementation and
additional numerical results can be found in Sec.~III of Ref.~\cite{Benic:2023ybl}.

\begin{figure}[htb]
  \begin{center}
  \includegraphics[scale = 0.8]{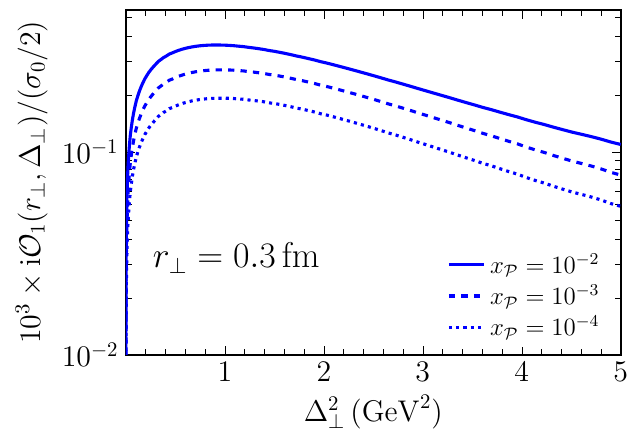}
  \end{center}
  \caption{The first azimuthal harmonic of the Odderon amplitude $\rmi\calO_1(r_\perp,\Delta_\perp)$ as a function of $\Delta_\perp^2$ for $r_\perp = 0.3$~fm and $\alpha_S=0.25$. Different line styles correspond to different values of $x_\calP$.
  }
  \label{fig:odder}
\end{figure}
Fig.~\ref{fig:odder} shows the first azimuthal harmonic of the Odderon amplitude $\rmi\calO_1(r_\perp,\Delta_\perp)$  as a function of
$\Delta_\perp^2$ for a dipole of size $r_\perp = 0.3$ fm. We have divided by the parameter $\sigma_0/2 = 16.36$ mb = 42.027 GeV$^{-2}$ (area of the proton) \cite{Lappi:2013zma} to compensate for the dimensionality of $\rmi\calO_1(r_\perp,\Delta_\perp)$. At $x_\calP=10^{-2}$ this corresponds to the
perturbative exchange of three gluons (with negative $C$-parity) while for smaller $x_\calP$ the BK resummation has been performed
as described above. We note, first of all, the small magnitude of $\calO_1(r_\perp,\Delta_\perp) < 10^{-2} \ll 1$, i.~e.~ 
that scattering of perturbative (small) dipoles via $C$-odd exchanges is predicted to be very weak. However, we also observe
that the $C$-odd amplitude peaks at fairly large $\Delta_\perp^2 \simeq 1$~GeV$^2$ and that it decreases by merely a factor of
$\sim 2$ as $\Delta_\perp^2$ increases further to $\simeq 3$~GeV$^2$. This is a manifestation of the ``Landshoff 
mechanism"~\cite{Landshoff:1974ew}, see also \cite{Donnachie:1979yu,Donnachie:1983ff,Donnachie:1983hf} whereby large-$t$ scattering via three gluon exchange
is less likely to break up the proton than scattering via two gluon exchange\footnote{The proton model light-cone wave
function employed here, however, includes a perturbative correction~\cite{Dumitru:2021tqp}
due to the emission of a gluon from one of the valence quarks, or the internal exchange of a gluon,
not considered in the quoted papers by Donnachie and Landshoff.}. In simple terms this is due to the fact
that a large momentum transfer can be shared by the exchanged gluons, and so a comparable transverse momentum can
be transmitted to up to three partons in the proton, resulting in a smaller increase of their invariant mass than in
case of two gluon exchange. Such a contribution has been illustrated in Fig.~\ref{fig:chicJ}. The light-cone Fock space amplitudes of the proton are strongly suppressed if
the invariant mass of the parton system is far from $\sim N_c \Lambda_\mathrm{QCD}$.
On the other hand, an amplitude which depends weakly on momentum transfer, in impact parameter space would peak
at small $b_\perp$. Amplitudes with negative parity must vanish at $b_\perp=0$, however, and hence are expected to have
small magnitude.

Fig.~\ref{fig:odder} also shows the decrease of the Odderon exchange amplitude with $x_\calP$ by approximately a factor of 2 from $x_\calP=10^{-2}$ to $x_\calP=10^{-4}$. The initial decrease of the Odderon amplitude has been understood to originate from gluon Regge trajectory suppressions in the BJKP equation \cite{Kovchegov:2003dm,Motyka:2005ep,Yao:2018vcg}. In App.~\ref{sec:linO} we have checked that by omitting the unitarity corrections, the Odderon still decreases but at a smaller rate. Importantly, the sign of the Odderon amplitude is preserved by the evolution equation, which is simply a consequence of the fact that its evolution equation is linear in $\calO(\rp,\bp)$.

\section{The Primakoff contribution}
\label{sec:prim}

In this section we focus on the Primakoff process, wherein the exclusive $\gamma^* p \to \calH p$ production proceeds via photon 
rather than Odderon exchange. Typically, the cross section for a QCD process far exceeds its QED counterparts, 
and so the latter can be safely neglected. 
This is, however, not so in case of the Odderon because its QCD cross section starts at order $\alpha_S^6$ and so the QED contribution 
becomes a competitive background in Odderon searches, in particular for low momentum transfer.

We treat the Primakoff process in the high-energy approximation of eikonal scattering, replacing the Odderon amplitude $\calO(\rp,\delp)$ in Eq.~\eqref{eq:M} with \cite{Dumitru:2019qec,Benic:2023ybl}
\be
\calO(\rp,\delp) \to 8\pi \rmi q_c \alpha \sin\left(\frac{\delp \cdot \rp}{2}\right)\frac{F_1(\Delta_\perp)}{\delp^2}\,,
\label{eq:photonXchangeAmpli}
\ee
at leading order in the electromagnetic coupling $\alpha = e^2/4\pi$.
The amplitude can be written as
\be
\left\langle\calM_{\lambda \lambdab}(\gamma^* p \to \calH p)\right\rangle \equiv -\frac{e F_1(\ell_\perp)}{\ellp^2} n_\mu\calM^\mu_{\lambda \lambdab}(\gamma^* \gamma^* \to \calH)\,,
\label{eq:2gamma}
\ee
where $\calM^\mu_{\lambda \lambdab}(\gamma^* \gamma^* \to \calH)$ is the $\gamma^*(\lambda,q)\gamma^*(\mu,\ell) \to \calH(\lambdab,\Delta)$ amplitude (in the high energy limit) that is obtained by 
inserting \eqref{eq:photonXchangeAmpli} into \eqref{eq:M}.
In \eqref{eq:2gamma}, $\ell \equiv P - P'$ is the momentum transfer from the proton, with $\ellp = 
\delp$ in the dipole frame and $F_1(\ell_\perp)$ is the Dirac charge form factor of the proton with $F_1(0) = 1$, see 
App.~\ref{sec:Odderon-sign} for more detail. Since \eqref{eq:photonXchangeAmpli} represents single photon exchange it is more convenient 
to start from the expression in Eq.~\eqref{eq:redA} for the reduced amplitude and perform the $\rp$ integral. This leads to
\be
\begin{split}
\frac{1}{q^-} n_\mu\calM^\mu_{\lambda \lambdab}(\gamma^*\gamma^* \to \calH) &= - 8\pi \rmi q_c^2 \alpha N_c \int_z \int_{\lp} \left[\frac{A_{\lambda\lambdab}(\lp - \bar{z}\delp,\lp,z)}{(\lp - \bar{z}\delp)^2 + \varepsilon^2} - \frac{A_{\lambda\lambdab}(\lp + z\delp,\lp,z)}{(\lp + z\delp)^2 + \varepsilon^2}\right]\frac{\phi_{\calH}(\lp,z)}{z\bar{z}}\,,
\end{split}
\label{eq:gen2}
\ee
The term $e F_1(\ell_\perp)/\ellp^2$ in Eq.~\eqref{eq:2gamma} is understood as a part of the usual Weizs\" acker-Williams photon field
\be
A^\mu(\ell) = (2\pi)\delta(\ell^-) \frac{1}{\ellp^2} e F_1(\ell_\perp)n^\mu\,,
\ee
where $n^\mu$ is the light-cone gauge vector.

\subsection{$|t| \to 0$ limit}

In the high energy approximation it is possible to express the $|t| \to 0$ limit of the Primakoff cross section 
for a $C$-even quarkonia $\calH$ with spin $J\neq 1$ in terms of the two-photon decay width $\Gamma(\calH\to\gamma\gamma)$
\be
\lim_{|t|\to 0} |t| \frac{\rmd\sigma(\gamma p \to \calH p)}{\rmd |t|} = \frac{8\pi \alpha (2J + 1) \Gamma(\calH \to \gamma\gamma)}{ M_{\calH}^3}\,.
\label{eq:m-general}
\ee
Eq.~\eqref{eq:m-general} is a model independent result and not affected by QCD corrections. Hence it represents a stringent constraint on the model predictions and a useful cross check. The key point for obtaining (\ref{eq:m-general}) is the relation of the high energy impact factor for the quarkonia photoproduction, 
 to the quarkonia's two-photon decay amplitude. 
We provide a derivation of (\ref{eq:m-general}) in App.~\ref{primakoff_limits}. For axial vector quarkonia, like $\chi_{c1}$, $\Gamma(\calA\to \gamma\gamma)$ vanishes and Eq.~(\ref{eq:m-general}) implies only that $\lim_{|t| \to 0}|t|\rmd\sigma(\gamma p \to \calA p)/ \rmd |t| = 0$, as is consistent with the Landau-Yang (LY) theorem \cite{Landau:1948kw,Yang:1950rg}, but does not constrain the value of the cross section at $|t| \to 0$. In App.~\ref{primakoff_limits} we thus provide a separate and general analysis of the $|t|\to 0$ limit for the axial quarkonia.

In practice, we will use \eqref{eq:m-general} to deduce $\Gamma(\calH \to \gamma\gamma)$ from the computation of its left hand side. This will be required in Sec.~\ref{sec:boosted} where we perform a fit of the quarkonia wave functions. Therefore, we now focus on obtaining explicit expressions of the $\gamma^* p \to \calH p$ amplitudes in the $|t| \to 0$ (and $Q^2 \to 0$) limit. 

Starting with the scalar quarkonia we use \eqref{eq:gen2} and insert the appropriate expressions for the reduced amplitudes $A_{\lambda,\lambdab}(\lp,\lonp,z)$ found in Sec.~\ref{sec:FinalAmplitudes}. 
The relevant reduced amplitude in the $Q^2 \to 0$ limit concerns only the second line of \eqref{eq:trscal}. The expression in the square bracket in \eqref{eq:gen2} is conveniently factored as
\be
\frac{A_{\lambda = \pm 1}(\lp - \bar{z}\delp,\lp,z)}{(\lp - \bar{z}\delp)^2 + \varepsilon^2} - \frac{A_{\lambda = \pm 1}(\lp + z\delp,\lp,z)}{(\lp + z\delp)^2 + \varepsilon^2} \to 2m_c \frac{(\epsp^\lambda\cdot\delp)}{(\lp^2 + m_c^2)^2}\left[\lp^2 + (z - \bar{z})^2 m_c^2\right]\,,
\label{eq:expandbrack}
\ee
where we have set $Q^2 = 0$, performed an expansion around $\Delta_\perp \to 0$ to linear order, as well as the angular average according to $l_\perp^i l_\perp^j \to \lp^2\delta^{ij}/2$.
We now insert \eqref{eq:expandbrack} into \eqref{eq:gen2} and use second line of \eqref{eq:Mscal} to extract $\wcalM_T$ as
\be
\wcalM_T \to 8\sqrt{2}\pi \rmi q_c^2 \alpha N_c m_c \frac{e F_1(\Delta_\perp)}{\Delta_\perp} \int_z\int_{\lp} \frac{\lp^2 + (z-\bar{z})^2 m_c^2}{(\lp^2 + m_c^2)^2}\frac{\phi_\calS(\lp,z)}{z\bar{z}}\,.
\label{eq:MSt0}
\ee

Performing similar steps for axial quarkonia we find that for the square brackets in \eqref{eq:gen2} the $O(\Delta_\perp)$ contributions vanish for $\lambdab = 0$ and for $\lambdab = \pm 1$. A non-vanishing contribution is found for $\lambdab = \pm 1$ at $O(\Delta_\perp^2)$ leading to
\be
\wcalM_{TT} \to - 16 \pi q_c^2 \alpha N_c eF_1(\Delta_\perp) \int_z \int_{\lp} \frac{(z-\bar{z})^2 m_c^2}{(\lp^2 + m_c^2)^2}\frac{\phi_\calA(\lp,z)}{z\bar{z}}\,,
\label{eq:MAt0}
\ee
while the $\lambdab = 0$ contribution is $O(\Delta_\perp^3)$. Such a special $\Delta_\perp$ dependence of the axial quarkonia amplitude is explained in App.~\ref{sec:axial} from general considerations.

For the calculation of the $|t|\to 0$ limit for the tensor quarkonia we also need 
\be
l_\perp^i l_\perp^j l_\perp^m l_\perp^n \to \frac{1}{8}\lp^4 (\delta^{ij}\delta^{mn} + \delta^{im}\delta^{jn} + \delta^{in}\delta^{jm})\,.
\ee
Computing for various tensor polarizations $\lambdab$ we find that $\lambdab = \pm 1$ does not contribute in the $|t| \to 0$ limit. As for the $\lambdab = \pm 2$ and $\lambdab = 0$ cases, we extract the amplitudes (\ref{eq:MpolTens}) as 
\be
\begin{split}
&\wcalM_{TT2,p} \to 16\sqrt{2}\pi\rmi q_c^2 \alpha N_c \frac{e F_1(\Delta_\perp)}{\Delta_\perp}\int_z \int_{\lp} \frac{\lp^2}{(\lp^2 + m_c^2)^2} \left[(z^2 +\bar{z}^2)\lp^2 + m_c^2\right]\frac{\phi_{\calT,T2}(\lp,z)}{z\bar{z}}\,,\\
&\wcalM_{TT2,f} \to 0\,,\\
&\wcalM_{TL} \to -\frac{16}{\sqrt{3}}\pi\rmi q_c^2 \alpha N_c \frac{e F_1(\Delta_\perp)}{\Delta_\perp} \int_z \int_{\lp} \frac{m_c^2}{(\lp^2 + m_c^2)^2} \left[\lp^2 - 2(z-\bar{z}^2)\left(\frac{3}{2}\lp^2 + m_c^2\right)\right] \frac{\phi_{\calT,L}(\lp,z)}{z\bar{z}}\,.
\end{split}
\label{eq:MTt0}
\ee

\subsection{Adding the Pauli form factor}
\label{sec:pauli}

In general, the $\gamma^* p \to \calH p$ amplitude can be related to the $\gamma^*\gamma^*\to \calH$ amplitude by separating out the Dirac current with the Dirac ($F_1$) and Pauli ($F_2$) form factors 
\be
\calM_{\lambda\lambdab} (\gamma^* p \to \calH p) = \calM_{\lambda\lambdab}^{\mu}(\gamma^*\gamma^* \to \calH) \frac{ g_{\mu\nu}}{\ell^2} \bar u_{h'}(P') \left[e F_1(\ell_\perp) \gamma^\nu + \rmi \sigma^{\nu\rho} \ell_\rho \frac{e F_2(\ell_\perp)}{2 m_N}\right] u_h(P)\,,
\label{eq:amppauli}
\ee
where $u_h(P)$ is a proton spinor with helicity $h$ and $\sigma^{\nu\rho} = \rmi[\gamma^\nu,\gamma^\rho]/2$. The dominant contribution of the first term in the high energy limit is with $\gamma^+$, while in the second term the dominant contribution is with $\sigma^{+ i}$. Thus, we can write in a covariant way 
\be
\calM_{\lambda\lambdab} (\gamma^* p \to \calH p) \approx P_\mu\calM_{\lambda\lambdab}^{\mu}(\gamma^*\gamma^* \to \calH)\frac{1}{q\cdot P} \bar u_{h'}(P') \left[\frac{e F_1(\ell_\perp)}{\ell^2} \slashed{q} + \rmi \sigma^{\nu\rho} q_\nu \ell_\rho \frac{e F_2(\ell_\perp)}{2 m_N \ell^2}\right] u_h(P)\,.
\label{eq:amppauli2}
\ee
The above spinor vertices in the high energy limit become\footnote{The calculation steps are very similar to the one performed in App.~\ref{sec:photonwf} and also available in \cite{Lepage:1980fj}.}
\be
\begin{split}
&\bar u_{h'}(P') \slashed{q} u_h(P) \approx W^2 \delta_{hh'}\,,\\
&\bar u_{h'}(P') \rmi \sigma^{\nu\rho} q_\nu \ell_\rho u_h(P) \approx  \ell_\perp h \rme^{\rmi h \phi_\ell} W^2 \delta_{h,-h'}\,.
\end{split}
\ee
With the relation  $\left\langle\calM_{\lambda\lambdab}\right\rangle = q^- \wcalM_{\lambda\lambdab} = q^- \calM_{\lambda\lambdab}/W^2$, we have
\be
\left\langle\calM_{\lambda\lambdab} (\gamma^* p \to \calH p)\right\rangle = - n_\mu \calM_{\lambda\lambdab}^{\mu}(\gamma^*\gamma^* \to \calH) \left[\frac{e F_1(\ell_\perp)}{\ellp^2}\delta_{hh'} + \frac{e F_2(\ell_\perp)}{\ellp^2}\frac{\ell_\perp}{2 m_N} h\rme^{\rmi h\phi_\ell}\delta_{h,-h'}\right]\,,
\label{eq:DirPau}
\ee
where the first term reproduces \eqref{eq:2gamma}. Thanks to the difference in the helicity structure in the high energy limit there is no interference between the $F_1$ and $F_2$ contributions in \eqref{eq:DirPau}. For the same reason there is no intereference of the $F_2$~term with the Odderon exchange amplitude. The addition of $F_2$ to the Primakoff cross section therefore amounts to a simple replacement
\be
F_1^2(\ell_\perp) \to F_1^2(\ell_\perp) + \frac{\ellp^2}{4 m_N^2} F_2^2(\ell_\perp)\,.
\ee
Thanks to the additional $\ell_\perp$ dependence, the $F_2$ contribution is negligible at $|t|  = \ellp^2 \to 0$ but in practice becomes up to about $50\%$ correction at finite $|t|$ \cite{Ye:2017gyb}.

\section{Boosted Gaussian model for the $\chi_{cJ}$ wave functions}
\label{sec:boosted}

In this section we perform a fit of the scalar part of the $C$-even quarkonia wave function $\phi_{\calH,i}(\kp,z)$. We adopt a Boosted Gaussian ansatz \cite{Kowalski:2006hc} originally written in coordinate space as
\be
\phi_{\calH,B}(r_\perp,z) = \calN_{\calH,B} z\bar{z} \exp\left(-\frac{m_c^2 \calR_\calH^2}{8 z\bar{z}} - \frac{2z\bar{z} r_\perp^2}{\calR_\calH^2} + \frac{1}{2}m_c^2 \calR_\calH^2\right)\,, 
\label{eq:bg}
\ee
where $\calR_\calH$ is the radii parameter and $\calN_{\calH,i}$ are normalization parameters for different quarkonia species and polarizations that are to be determined below. For $\calH = \calS$ we have only one normalization, $\calN_\calS$, for $\calH = \calA$, we have $B = T$, $L$, denoting transverse and longitudinal polarizations. For $\calH = \calT$, the index $B$ spans over $B = T2$, $T$, $L$.
The wave function is normalized as
\be
1 = N_c \sum_{h\bar{h}}\int_z \int_{\rp} \left|\Psi^\calH_{\lambdab,h\bar{h}}(\rp,z)\right|^2\,.
\label{eq:normft}
\ee
In practice, we convert \eqref{eq:normft} to momentum space and express the helicity sum as a Dirac trace as
\be
1 = N_c \int_{\kp}\int_z \frac{1}{z\bar{z}}{\rm tr}\left[(\slashed{k} - m_c)\gamma^0 \Gamma^{\calH,\dag}_{\lambdab}(k,k')\gamma^0 (\slashed{k} + m_c)\Gamma^{\calH}_{\lambdab}(k,k')\right]\left|\phi_{\calH,B}(\kp,z)\right|^2\,.
\label{eq:normdir}
\ee

To constrain the parameters $\calR_\calH$ and $\calN_{\calH,i}$, we use \eqref{eq:normdir} and the $\calH \to \gamma\gamma$ decay width for $\chi_{c0}$ and $\chi_{c2}$. The $\chi_{c1}\to\gamma\gamma$ decay is forbidden due to the LY theorem. In this case our assumption is that the $\calR$ parameter of $\chi_{c1}$ is equal to that for $\chi_{c2}$.

The decay rate for $\calH = \calS,$ $\calT$ can be obtained from the correspondence in Eq.~\eqref{eq:m-general}. For scalar quarkonia we first insert the amplitude \eqref{eq:MSt0} into \eqref{eq:csS} to find the $|t|\to 0$ limit of the cross section. We then extract the decay rate via \eqref{eq:m-general} as
\be
\begin{split}
& \Gamma(\calS \to \gamma\gamma) = \frac{\pi \alpha^2}{4} M_\calS^3 F_\calS^2\,,\\
& F_\calS \equiv 4 q_c^2 m_c N_c\int_z\int_{\kp} \frac{\kp^2 + (z-\bar{z})^2 m_c^2}{(\kp^2 + m_c^2)^2}\frac{\phi_\calS(\kp,z)}{z\bar{z}}\,,
\label{eq:GAS}
\end{split}
\ee
which agrees perfectly with Eq.~(13) in Ref.~\cite{Li:2021ejv} after adjusting the conventions. We have additionally confirmed that in the NRQCD limit \eqref{eq:GAS} agrees with the known result in \cite{Kwong:1987ak} (see App.~\ref{primakoff_limits}).


For the actual fit, the LO result in \eqref{eq:GAS} is supplemented by the NLO QCD corrections \cite{Barbieri:1980yp,Kwong:1987ak}. This amounts to a replacement 
\be
\Gamma(\calS \to \gamma\gamma) \to \left(1 + \frac{\alpha_S}{\pi}\left(\frac{\pi^2}{3} - \frac{28}{9}\right)\right)\Gamma(\calS \to \gamma\gamma) \approx (1 + 0.06\alpha_S)\Gamma(\calS \to \gamma\gamma)\,.
\ee
For the experimental value of $\chi_{c0} \to \gamma\gamma$ decay width we use the most recent PDG value: $\Gamma(\chi_{c0}\to\gamma\gamma) = 2.203\times 10^{-6}$~GeV~\cite{ParticleDataGroup:2022pth}.
Using $m_c = 1.4$ GeV and $M_{\chi_{c0}} = 3.414$~GeV and $\alpha_S(2m_c) \approx 0.25$, 
we determine $\calN_{\chi_{c0}} = 1.148$ and $\calR_{\chi_{c0}} = 1.539$ GeV$^{-1}$. As expected, the radius parameter $\calR$  of
the $\chi_{c0}$ is similar to that of the $J/\psi$~\cite{Kowalski:2006hc}.

For $\calT$ we insert the amplitude \eqref{eq:MTt0} into \eqref{eq:csT} and extract the decay rate via \eqref{eq:m-general} as
\be
\begin{split}
& \Gamma(\calT \to \gamma\gamma) = \frac{\pi \alpha^2}{20} M_\calT^3 \left(F_{\calT,L}^2 + F_{\calT,T2}^2\right)\,,\\
& F_{\calT,L} = \frac{4\sqrt{2}}{\sqrt{3}} q_c^2 N_c \int_z \int_{\kp}\frac{m_c^2\left[\kp^2 - 2(z-\bar{z})^2 \left(\frac{3}{2}\kp^2 +  m_c^2\right)\right]}{(\kp^2 + m_c^2)^2}\frac{\phi_{\calT,L}(\kp,z)}{z\bar{z}}\,,\\
& F_{\calT,T2} = 8 q_c^2 N_c \int_z \int_{\kp}\frac{\kp^2\left[(z^2 + \bar{z}^2)\kp^2 + m_c^2\right]}{(\kp^2 + m_c^2)^2}\frac{\phi_{\calT,T2}(\kp,z)}{z\bar{z}}\,,
\end{split}
\label{eq:GAT}
\ee
where $F_{\calT,L}$ ($F_{\calT,T2}$) originates from the $\lambdab = 0$ ($\lambdab = \pm 2$) polarizations. We find only a partial agreement of our result \eqref{eq:GAT} and Eqs.~(22)-(24) of \cite{Li:2021ejv}. The third line of \eqref{eq:GAT} can be brought in agreement with Eq.~(24) in \cite{Li:2021ejv} after appropriate adjustments in conventions and also after a judicious identification of $M_\calT$ with the pair invariant mass $M_0$. However, the square brackets of the second line in \eqref{eq:GAT} contain a term $\kp^2$ while in (23) of \cite{Li:2021ejv} they rather have $-m_c^2$. Using \eqref{eq:GAT} in the NRQCD limit we find the $\lambdab = 0$ contribution vanishes with the $\lambdab = \pm 2$ contribution saturating the NRQCD limit completely and in accordance with the known result \cite{Kwong:1987ak}. Taking the NRQCD limit of (22)-(24) in \cite{Li:2021ejv} gives an incorrect result as it leads to a finite contribution from the $\lambdab = 0$ polarization.

The QCD corrections \cite{Barbieri:1980yp,Kwong:1987ak} turn out to be sizeable in this particular channel \cite{Barbieri:1980yp,Kwong:1987ak}
\be
\Gamma(\calT \to \gamma\gamma) \to \left(1 - \frac{16\alpha_S}{3\pi}\right)\Gamma(\calT \to \gamma\gamma) \approx (1 - 1.7\alpha_S)\Gamma(\calT \to \gamma\gamma)\,.
\ee
Using $M_{\chi_{c2}} = 3.556$~GeV and the PDG value $\Gamma(\chi_{c2}\to\gamma\gamma) = 5.614\times 10^{-7}$ GeV~\cite{ParticleDataGroup:2022pth}, we find $\calN_{\chi_{c2},T2} = 0.609$ GeV$^{-1}$ $ = \calN_{\chi_{c2},L}$, $\calN_{\chi_{c2},T} = 0.592$ GeV$^{-1}$ and $\calR_{\chi_{c2}} = 1.482$ GeV$^{-1}$. In these numerical estimates we used $\alpha_S = 0.25$ in the NLO correction factors.

Finally, for $\chi_{c1}$ we assume $\calR_{\chi_{c1}} = \calR_{\chi_{c2}}$, allowing us to fix the parameters $\calN_{\chi_{c1},T,L}$ via normalization. We find $\calN_{\chi_{c1},T} = 1.388$ and $\calN_{\chi_{c1},L} = 1.401$. 

From the above fits we may compute charge radii of the $\chi_{cJ}$ states via the following formula~\cite{Li:2017mlw}:
\be
\left\langle r_c^2 \right\rangle = N_c\sum_{h\bar{h}}\int_z \int_{\rp} 
\left(A_{\lambdab}^{\calH} \bar{z}^2 \rp^2\right) \left|\Psi^\calH_{\lambdab,h\bar{h}}(\rp,z)\right|^2\,.
\ee
$(A_{\lambdab}^{\calH})^{-1}$ denotes the expectation value of $\rp^2/\boldsymbol{r}^2
= \sin^2\theta$ in the state
\be
|J, \lambdab\rangle = \sum_{m,m_s} C(J,\lambdab| L,m; S,m_s) \, |L,m\rangle\, |S,m_s\rangle~,
\ee
where $C(J,\lambdab| L,m; S,m_s)$ denotes the Clebsch-Gordan coefficient for
$L=S=1$. This gives $A^{\chi_{c0}}=3/2$, $A^{\chi_{c1}}_0=5/4$, 
$A^{\chi_{c1}}_1=5/3$, $A^{\chi_{c2}}_0=15/8$, $A^{\chi_{c2}}_1=5/3$, $A^{\chi_{c2}}_2=5/4$.
For $\chi_{c0}$ we then find $\sqrt{\langle r^2_c\rangle} = 0.270$ fm, which is close to the value obtained in Ref.~\cite{Li:2017mlw} from potential models. 
For $\chi_{c1}$ we find $\sqrt{\langle r^2_c\rangle} = 0.277$ fm for $\lambdab = \pm 1$, while $\sqrt{\langle r^2_c\rangle} = 0.236$ fm was obtained for $\lambdab = 0$. For $\chi_{c2}$ we have $\sqrt{\langle r^2_c\rangle} = 0.263$ fm for $\lambdab = \pm 2$, $\sqrt{\langle r^2_c\rangle} = 0.262$ fm for $\lambdab = \pm 1$ and $\sqrt{\langle r^2_c\rangle} = 0.237$ fm for $\lambdab = 0$. 
Thus, all charge radii are of comparable magnitude and less than 0.3~fm, which appears reasonable.

\section{Numerical results}
\label{sec:numer}

In this Section we will first show numerical results for the $\gamma^* p \to \chi_{cJ} p$ cross section based on the formulas in Secs.~\ref{sec:FinalAmplitudes} and \ref{sec:prim}, for EIC kinematics. Using the $\gamma^* p \to \chi_{cJ} p$ cross sections, we will also compute the electro-production cross section $e(k) p(P) \to \chi_{cJ}(\Delta) e(k') p(P')$ through the standard formula~\cite{vonWeizsacker:1934nji, Williams:1935dka, Budnev:1975poe, Caldwell:2009ke}
\be
\frac{\rmd \sigma_{ep}}{\rmd x_\calP \rmd Q^2 \rmd |t|} = \frac{\alpha}{2\pi Q^2 x_\calP}\left\{2(1-y) \frac{\rmd \sigma_L}{\rmd |t|} + \left(1 + (1-y)^2 - 2(1-y)\frac{Q^2_{\rm min}}{Q^2}\right)\frac{\rmd \sigma_T}{\rmd |t|} \right\}\,,
\label{eq:electro1}
\ee
where $Q^2_{\rm min} = m_e^2 y^2/(1-y)$ accounts for projectile mass corrections, with $m_e$ being the electron mass. $y$ is the inelasticity,
given by
\be
y \equiv \frac{q\cdot P}{k\cdot P}  = \frac{W^2 + Q^2 - m_N^2}{S - m_N^2}\,,
\ee
where $S \equiv (k + P)^2$ is the $ep$ collision energy, $W^2 \equiv (q + P)^2$ is the $\gamma^* p$ collision energy and $m_N$ is the proton mass. Recall that the Odderon amplitude evolves with Pomeron-$x_\calP$, which is related to the conventional Bjorken-$x$ variable $x_B = Q^2/(2 q\cdot P) = Q^2/(W^2 + Q^2 - m_N^2)$ via 
\be
x_\calP\equiv \frac{x_B}{\beta}\,,\qquad \beta \equiv \frac{Q^2}{2q\cdot\ell} = \frac{Q^2}{Q^2 + M_\calH^2 + |t|}\,.
\label{eq:pomx}
\ee

In the numerical computations, we augment the Primakoff cross section by taking into account the Pauli form factor $F_2$ as explained in Sec.~\ref{sec:pauli}. For $\chi_{c0}$ and $\chi_{c2}$ we take into account the available QCD corrections to the Primakoff cross section via \eqref{eq:m-general}. The total cross section is based on taking the coherent sum of the Primakoff and the Odderon exchange amplitudes in which case their relative sign becomes crucial. Our careful analysis in App.~\ref{sec:Odderon-sign} reveals that the {\it Primakoff and Odderon amplitudes are in-phase}, that is, {\it they interfere constructively} for each value of the Odderon evolution parameter $x_{\calP}$, as we have also explained in Sec.~\ref{sec:Odderon-Evol}. The contribution of $F_2$ does not interfere with the coherent sum of the Odderon and Primakoff amplitudes. The interference between the QCD correction to the Primakoff amplitude (available for $\chi_{c0,2}$) and the Odderon exchange is determined through the relative phase of the Odderon and the Primakoff amplitude at tree level. For the Odderon component we use the solutions of the rcBK evolution, keeping only the $k = 0$ harmonic of the Fourier series \eqref{eq:omom}, as explained in Sec.~\ref{sec:Odderon-Evol}. For the Primakoff component we use the recent fits of $F_1$ and $F_2$ from \cite{Ye:2017gyb}. Our standard choice for the QCD coupling $\alpha_S$ is  $\alpha_S(2 m_c) \approx 0.25$ unless stated otherwise.
\\
 
\begin{figure}[htb]
  \begin{center}
  \includegraphics[scale = 0.7]{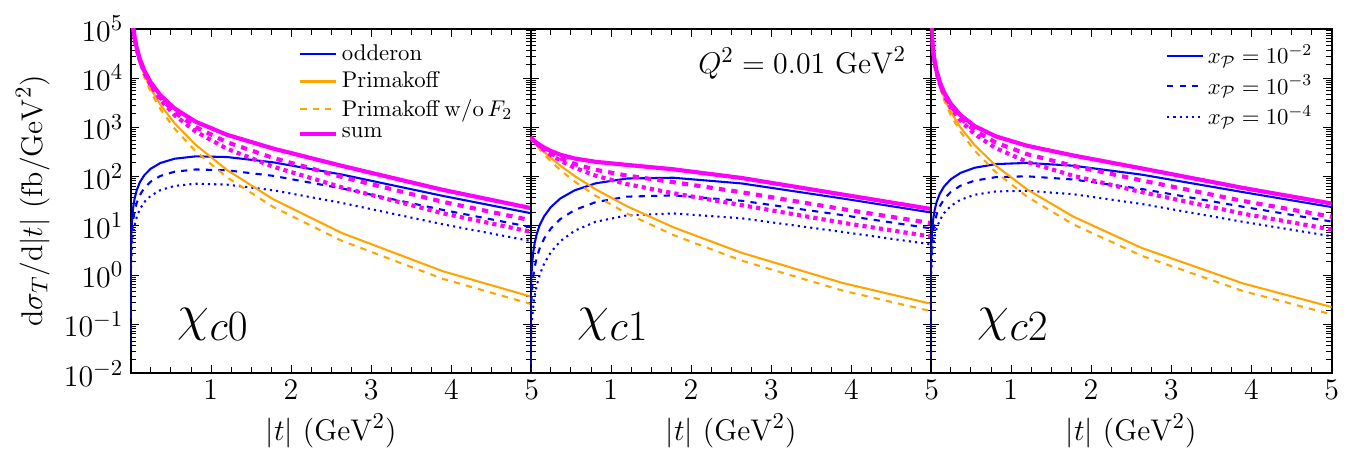}
  \end{center} \vspace*{-0.5cm}
  \caption{The $\gamma^* p \to \chi_{cJ} p$ cross sections as functions of $|t|$ for the transverse photon. The label ``sum" stands for the coherent sum of the odderon and the Primakoff contributions. We have set $Q^2 = 0.01$ GeV$^2$. Different line styles in the odderon and the summed cross section correspond to different values of $x_\calP$. The comparison of the Primakoff contribution with the full and dashed orange lines shows the impact of $F_2$.}
  \label{fig:csTL}
\end{figure}
We start with the numerical results for the $\gamma^* p \to \chi_{cJ} p$ cross sections shown in Fig.~\ref{fig:csTL} as functions of $|t|$, 
for different values of $x_\calP$. We have set $Q^2 = 0.01$ GeV$^2$. Since $Q^2$ is low, we focus on the transverse cross section $\rmd\sigma_T/\rmd |t|$.
At small $|t|$ the cross section is, of course, dominated by the Primakoff process (photon exchange).
However, we note that our predictions for the Primakoff cross sections are substantially lower than those
shown in Fig.~4c of Ref.~\cite{Jia:2022oyl}. This emphasizes the importance of the constraint~(\ref{eq:m-general})
from the two photon decay rate.

The QCD Odderon exchange amplitude reaches a comparable magnitude at $|t|\approx 1$~GeV$^2$, depending on $x_\calP$ and $J$. 
The lowest cross-over from the Primakoff dominated to the Odderon dominated regime at $|t| < 1$~GeV$^2$ is seen for $\chi_{c1}$ and $\chi_{c2}$ where the Primakoff-like background is lower than for $\chi_{c0}$. In the regime of $|t|$ where the individual Primakoff and Odderon cross sections are of similar magnitude, thanks to their constructive interference, the coherently summed cross section is four times greater than the Primakoff cross section alone.
At high momentum transfer Odderon exchange dominates due to its slower fall-off with $|t|$.

\begin{figure}[htb]
  \begin{center}
  \includegraphics[scale = 0.7]{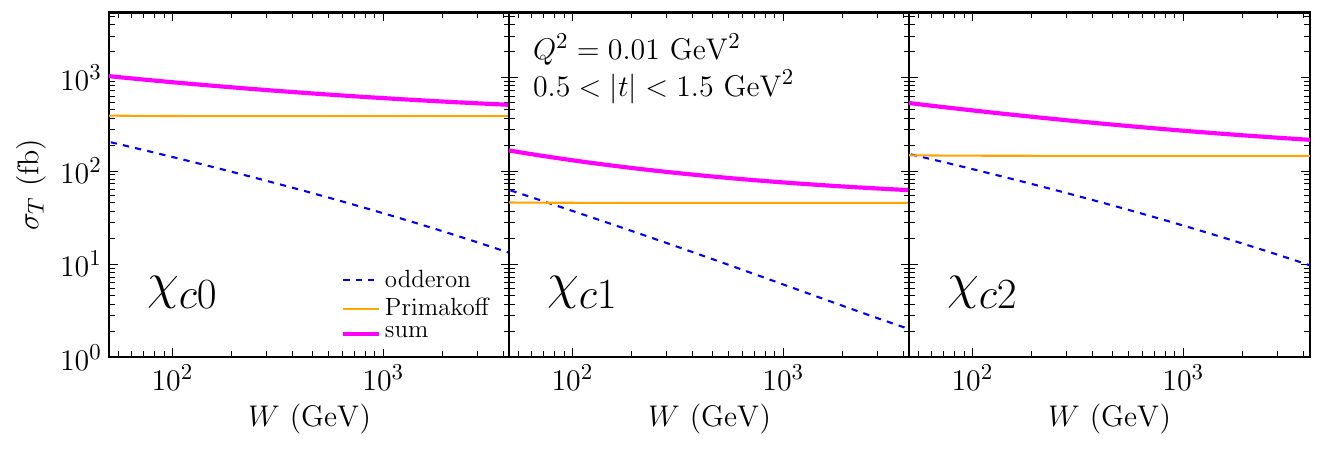}
  \end{center} \vspace*{-0.5cm}
  \caption{The $\gamma^* p \to \chi_{cJ} p$ cross sections as functions of $W$ for the transverse photon. The label ``sum" stands for the coherent sum of the Odderon and Primakoff contributions. We have set $Q^2 = 0.01$ GeV$^2$ and integrated in $|t|$
  over $0.5 < |t| < 1.5$ GeV$^2$.}
\label{fig:csTW}
\end{figure}
For Fig.~\ref{fig:csTW} we have integrated $\rmd \sigma_T / \rmd |t|$ over the range $0.5 < |t| < 1.5$ GeV$^2$
as appropriate for the EIC design detector~\cite{AbdulKhalek:2021gbh}. We show the result as a function of $W$. As a consequence of small-$x$ evolution, the Odderon cross section drops with increasing $W$. 
As the Primakoff cross section is constant, at the
lower end of $W$ the coherent sum 
is about five times greater than the Primakoff component alone. 
Interestingly, $\sigma_T$ displays a
negative slope 
due to the decreasing Odderon amplitude towards smaller $x$.

 We find that the decrease of the Odderon cross section with $W$ is driven mostly by the nonlinear corrections in the unitarized evolution for the Odderon. In App.~\ref{sec:linO} (see Fig.~\ref{fig:bfklbk}, right) we have computed $\sigma_T$ based on linear evolution of the BKP Odderon and found a slower decrease with $W$ as anticipated from the asymptotics of the BLV solution with an intercept equal to one \cite{Bartels:1999yt}.
 
\begin{figure}[htb]
  \begin{center}
  \includegraphics[scale = 0.7]{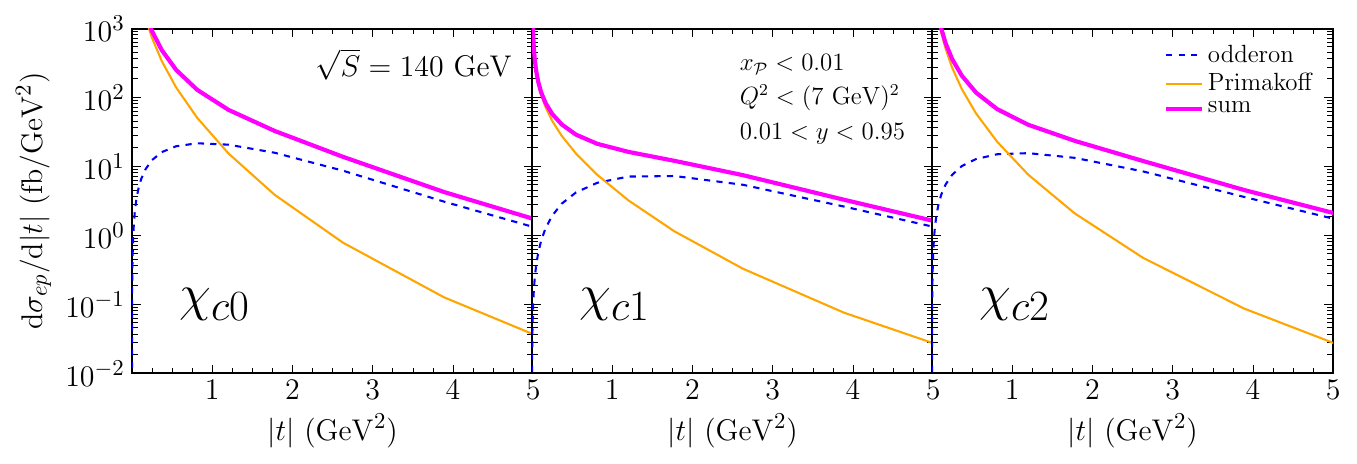}
  \end{center} \vspace*{-0.5cm}
  \caption{The $e p \to \chi_{cJ} e p$ ($J = 0$, $1$, $2$) cross sections as functions of $|t|$ at the top EIC energy of $\sqrt{S} = 140$ GeV.}
  \label{fig:cselec}
\end{figure}
We also calculate the differential $e p \to \chi_{c J}ep$ cross section for top EIC energy, $\sqrt{S} = 140$~GeV, and use the following kinematic cuts: $0.01 < y < 0.95$ \cite{AbdulKhalek:2021gbh}, $x_{\cal P} < 0.01$, and $Q^2_{\rm min}<Q^2 < (2\times 3.5 \, {\rm GeV})^2$. The resulting $|t|$-dependence of the cross section is shown in Fig.~\ref{fig:cselec}. With these cuts the crossover from photon to Odderon exchange
occurs at about $|t|\approx 1$~GeV$^2$ where the total cross section is several times greater than due to the Primakoff process alone.

\begin{figure}[htb]
  \begin{center}
  \includegraphics[scale = 0.7]{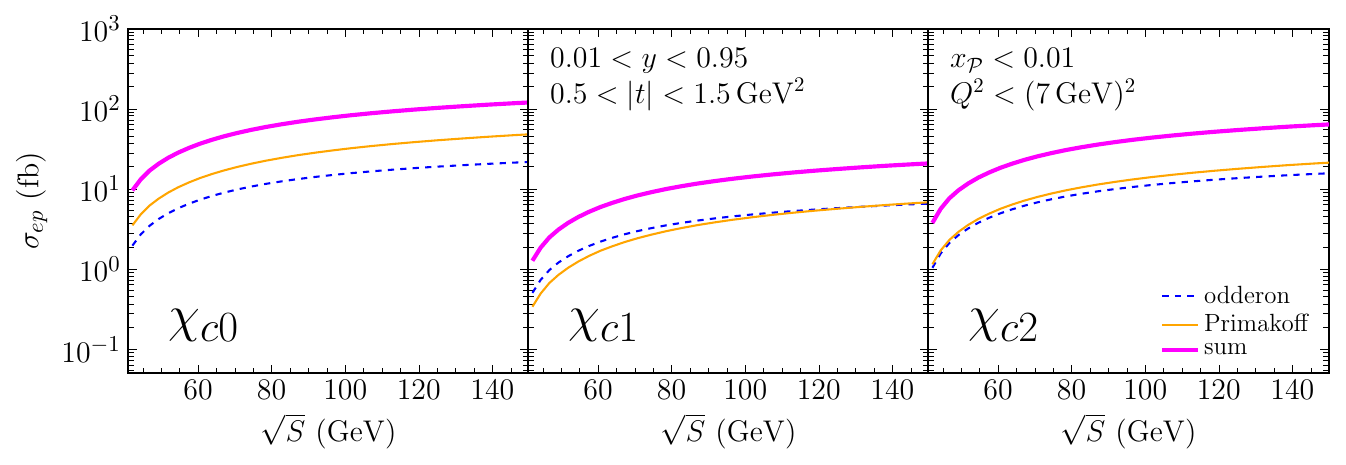}
  \end{center} \vspace*{-0.5cm}
  \caption{Total electroproduction cross section of the $\chi_{cJ}$ quarkonia as a function of the $ep$ center of mass energy $\sqrt{S}$. Kinematic cuts are explained in the text. 
  }
  \label{fig:tot}
\end{figure}
The total $\chi_{cJ}$ cross sections are shown in Fig.~\ref{fig:tot}. They have been
integrated over $|t|$ over the range $0.5$~GeV$^2 < |t| < 1.5$~GeV$^2$ where 
according to Fig.~\ref{fig:cselec} the Odderon contribution is appreciable. 
Both the photon and Odderon exchange contributions level off towards top EIC energy since our kinematic cuts
are energy independent, and neither exchange involves a positive intercept.

$\chi_{c1}$ has the highest branching ratio ${\rm BR}(\chi_{c1} \to J/\psi \gamma) = 34.3\%$ \cite{ParticleDataGroup:2022pth}.
As an example, let us estimate the total number of $\chi_{c1}$s per month at the EIC. From Fig.~\ref{fig:tot} the total cross section is about $\sigma_{ep} \approx 20$~fb at $\sqrt{S} = 140$~GeV. Multiplying by the
expected luminosity at the EIC ($L = 10^{34}$ cm$^{-2}$s$^{-1}=10^{-5}$ fb$^{-1}$ s$^{-1}$) gives
about $2\times 10^{-4}$ events/second or $\approx 518$ events/month. After taking into account ${\rm BR}(\chi_{c1} \to J/\psi \gamma)$ results in about 177 events/month. 
$J/\psi$'s are detected through the $J/\psi \to e^+ e^-$ or $\to \mu^+ \mu^-$ decays, and the combined
corresponding branching ratio is about ${\rm BR}(J/\psi\to \ell^+ \ell^-) = 12\%$, after which we end up with about 21 events/month.

\begin{figure}[htb]
  \begin{center}
  \includegraphics[scale = 0.7]{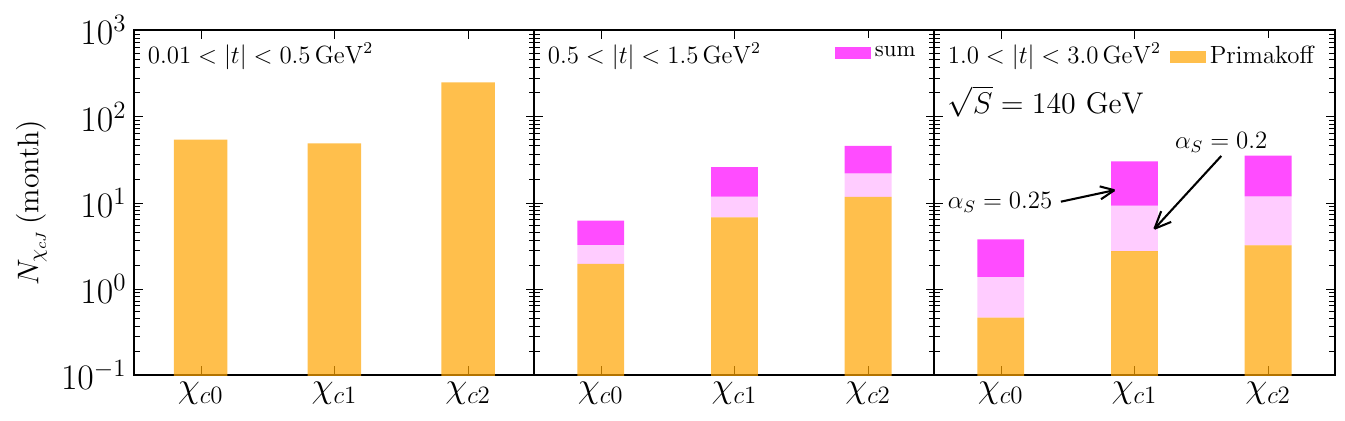}
  \end{center}
  \caption{A summary of expected Primakoff (orange) vs.\ total (magenta) number of exclusive $\chi_{cJ} \to
  J/\psi \gamma \to l^+ l^- \gamma$ events per month at the top EIC energy of $\sqrt{S} = 140$ GeV, for three different $|t|$ cuts. The remaining kinematic cuts are the same as in Fig.~\ref{fig:tot}. The two different bins for the total number of events 
  correspond to $\alpha_S = 0.25$ and $\alpha_S = 0.2$, respectively.}
  \label{fig:events}
\end{figure}
In the final Fig.~\ref{fig:events}, we summarize the expected number of exclusive $\chi_{cJ}$  events per month, $N_{\chi_{cJ}}$, at the EIC design luminosity
for two ranges of momentum transfer $|t|$. The result is obtained according to:
\be
N_{\chi_{cJ}} = L\times \sigma_{ep}(ep\to\chi_{cJ}ep)\times {\rm BR}(\chi_{cJ}\to J/\psi \gamma)\times {\rm BR}(J/\psi\to \ell^+ \ell^-)\,.
\ee
The most statistics is expected for $\chi_{c1}$ and $\chi_{c2}$ tensor quarkonia with about a factor of 2-3 excess events over the Primakoff process in the interval $0.5$~GeV$^2<|t|<1.5$~GeV$^2$. For $\chi_{c0}$, which has the largest cross section from Fig.~\ref{fig:tot} is compensated by a small branching ratio ${\rm BR}(\chi_{c0}\to J/\psi\gamma) = 1.4\%$ \cite{ParticleDataGroup:2022pth}. 

The presented estimates of $\chi_{cJ}$ production cross sections carry some theoretical uncertainties. For the photon exchange contribution they are smaller than for the Odderon exchange. The coupling of the photon to the proton is well constrained experimentally \cite{Ye:2017gyb}, and so the main sources of uncertainty are the $\gamma^*\gamma^* \to \chi_{cJ}$ amplitudes. They depend on the details of the quarkonia wave functions and are sensitive to unknown higher order QCD corrections. For $\chi_{c0}$ and $\chi_{c2}$ the differential cross section $\rmd\sigma/\rmd|t|$ for the Primakoff process obeys a stringent constraint at small $|t| \ll 1$~GeV$^2$ imposed by Eq.~(\ref{eq:m-general}) that holds at all orders in QCD. Hence, for these charmonia the uncertainties mentioned above affect mostly the details of the $t$-shape of $\rmd\sigma/\rmd|t|$. 
They are expected to be small as the $\chi_{cJ}$ wave functions are probed mostly at short distances $\sim 1/m_c$, where they are well constrained by the $\gamma\gamma$ decay width. The value of the coupling $\alpha_S(2m_c) \simeq 0.25$ is not large and the $t \to 0$ limit of the QCD corrections is known, so the uncertainties from QCD corrections at $t \neq 0$ should be small as well. For $\chi_{c1}$ the theoretical uncertainty of the Primakoff cross section is larger than for $\chi_{c0}$ and $\chi_{c2}$, as the constraint (\ref{eq:m-general}) cannot be imposed due to the LY theorem and, in consequence, the vanishing $\gamma\gamma$ decay width of $\chi_{c1}$.
In this case measurements of the differential cross sections $\rmd\sigma / \rmd|t|$ for $|t| \ll 1$~GeV$^2$ where the photon exchange dominates should greatly reduce the uncertainty associated with this contribution at $|t| \sim 1$~GeV$^2$, where we expect to isolate a significant Odderon signal.

In the Odderon exchange, in addition to uncertainties from unknown details of the quarkonia wave function and from higher order QCD corrections, there are uncertainties associated with the model for the proton wave function, and from the value of $\alpha_S$. The model of the proton employed here obeys general constraints coming from the measurements of the proton size, exclusive $J/\psi$ production off the proton~\cite{Dumitru:2019qec} and open charm electroproduction at HERA~\cite{Dumitru:2022ooz}. Those constraints, however, originate from measurements in the $C$-even sector, and the emerging $C$-odd correlators have never been probed experimentally. Furthermore, the perturbative Odderon is strongly sensitive to the numerical value of $\alpha_S$ -- the amplitude being $O(\alpha_S^3)$ at the lowest order. To quantify the uncertainty in the value of $\alpha_S$ we take an interval $0.2 < \alpha_S < 0.25$. The lower value $\alpha_S=0.2$ comes from a fit of the proton and $C$-even exchange models to open charm electroproduction at HERA~\cite{Dumitru:2022ooz}. Therein the average value of the hard scale $\mu = \sqrt{Q^2 + 4m_c^2}$ is about $5$~GeV, so the obtained value $\alpha_S =0.2$ is fully consistent with the running of $\alpha_S(\mu)$. In our case the bulk of exclusive $\chi_{cJ}$ production in $ep$ collisions occurs at small photon virtualities and moderate $|t|$, and so $\alpha_S(2 m_c) \approx 0.25$ is more appropriate. The range $0.2<\alpha_S<0.25$ results in an uncertainty of about a factor of 2.5 between the minimal and maximal values of the Odderon amplitude, when NLO effects in the proton wave function are included.

For the combined Primakoff and Odderon contributions, the $\alpha_S$-induced uncertainty is negligible for $|t| \ll 1$~GeV$^2$, where the Primakoff process dominates; whereas it about a factor of~$2$ when Primakoff and Odderon (with $\alpha_S =0.25$) amplitudes are close to each other, and up to a factor of about~$6$ when the Odderon strongly dominates over the Primakoff channel. Since the uncertainty coming from $\alpha_S$ is sizeable, we assume that it is the dominant uncertainty for the Odderon exchange and take it as our estimate of the theoretical uncertainty. We consider the choice of $\alpha_S = 0.25$ to be realistic, as the scale $2m_c$ is already fairly high. Another choice is $\mu = m_c$, leading to $\alpha_S \approx 0.35$, 
as obtained for exclusive $J/\psi$ photoproduction using the same model for the proton at leading order \cite{Dumitru:2019qec}.
Thus, we consider the lower value of $\alpha_S=0.2$ to be a conservative choice.

\section{Summary and Conclusions}
\label{sec:sum}

In this paper we have 
derived amplitudes for exclusive scalar, axial-vector and tensor quarkonium production,
Eqs.~(\ref{eq:Ampscal}, \ref{eq:Mpolscal}), (\ref{eq:Amps}, \ref{eq:Mpol}),
and (\ref{eq:redcT2}, \ref{eq:MpolTens}), respectively, in electron-proton scattering.
We have
provided first estimates of the cross sections for exclusive production
of $\chi_{cJ}$ quarkonia with positive $C$-parity at the EIC.
This process
requires a $C$-odd exchange in the $t$-channel. In the limit of heavy quarks and/or high
transverse momentum transfer this could be the exchange of a photon,
a Primakoff process, or the exchange of a color-symmetric three gluon ladder, the Odderon.  Our
estimates suggest that for $|t|\approx 1$~GeV$^2$ the two amplitudes are of similar magnitude
and that there are strong interference effects. Importantly, the relative phase
is not affected by QCD evolution of the Odderon towards small $x$, and so it is
determined by the three gluon exchange amplitude at moderately small $x_0$.
In turn, the matrix element of the eikonal color current operator $d^{abc} J^{+a} J^{+b} J^{+c}$
at $x_0$, for transverse momenta of order $\mathrm{max}(R^{-1}_{\chi_c}, |t|^{1/2})$ ($R_{\chi_c}$ being here roughly the size of $\chi_c$),
can be computed in a truncated Fock space for the proton that encompasses
the states that are relevant in that kinematic regime.

We find that photon and Odderon exchanges interfere constructively, leading to an
enhancement of the differential cross section for $\chi_{cJ}$ production around $|t|\approx 1$~GeV$^2$
by up to a factor of 4 over the pure photon exchange contribution.
Given that both the normalization and the $t$-dependence of the Primakoff
process are reasonably well determined, this presents a very exciting opportunity
to potentially discover at the EIC the hard $C$-odd exchange predicted by QCD.
Furthermore, we find that towards top EIC energy, $\sqrt S = 140$~GeV,
the total electroproduction cross section of $\chi_{cJ}$ quarkonia
(with kinematic cuts specified in the previous section) levels off
as neither photon nor Odderon exchange involve a positive intercept.
Therefore, it would be
important to measure the energy dependence~\cite{Aschenauer:2017jsk} 
of the cross section from $\sqrt S = 50$~GeV up to top EIC energy $\sqrt S = 140$~GeV,
where photon and Odderon amplitudes are similar, and where constructive
interference of amplitudes leads to a total cross section for production of
$\chi_{c1}$ and $\chi_{c2}$ quarkonia which exceeds the Primakoff component by a factor of $\approx 3-4$.

The predictions for the Odderon exchange process of course involve a number of
uncertainties such as the matrix element in the proton of $d^{abc} J^{+a} J^{+b} J^{+c}$
or the value of the strong coupling $\alpha_S$. The associated cross section scales
approximately like $\alpha_S^6$. 
Most importantly, the discovery
of the hard QCD $C$-odd exchange requires fixing the normalization of the
Primakoff background from measurements at low $\sqrt{|t|}< 0.5$~GeV; the relation~(\ref{eq:m-general})
of the $t\to0$ limit of the cross section to the $\gamma\gamma$ decay width provides
an important constraint on theoretical predictions for this background. The $t$-dependence
of the Primakoff cross section is then determined by the Dirac and Pauli electromagnetic form factors
of the proton which are well known.
Hence, one may then look for excess events
above this background, and for a change of slope of the differential cross section
$\mathrm{d}\sigma/\mathrm{d}t$, at higher $|t|\sim 1$~GeV$^2$ and beyond.

Our estimates indicate very weak dipole-proton (hard) scattering with $C$-odd exchange;
this is largely since these amplitudes are also parity odd and vanish for
impact parameter $\bp\to 0$, unlike parity even amplitudes, as well as due to the fact
that they fall off more rapidly towards large $\bp$.
The total electro-production cross section of $\chi_{cJ}$ quarkonia at top EIC energy
is estimated to have a magnitude $10-20$~fb (for $J=1$) and $60$~fb (for $J=2$).
Thus, it was not possible to observe these processes at the HERA accelerator.
However, in view of the projected high luminosity of the EIC, data collection over
a time span of several months to a year may be sufficient for the discovery of the
hard Odderon. A promising alternative would be to allow low-mass excitations of the proton, while requiring
a large rapidity gap to the $C=+1$ quarkonia. Such rapidity gap, diffractive processes have greater
cross-sections than exclusive ones.  Furthermore, they would extend the reach to higher $|t|$ where
Odderon exchange would more clearly dominate over photon exchange.


\begin{acknowledgments}
We thank Y.~Hatta, H.~M\"antysaari, L.~Pentchev, and R.~Venugopalan for useful discussions.
The work of S.~B. and A.~K. is supported by the Croatian Science Foundation (HRZZ) no. 5332 (UIP-2019-04).
A.~D. acknowledges support by the DOE Office of Nuclear Physics through
Grant DE-SC0002307, and The City University of New York for PSC-CUNY
Research grant 65079-00 53. He also thanks the EIC Theory Institute at
Brookhaven National Laboratory for their hospitality during a visit
in July 2023 when initial work on this project was done. 
L.~M.\ acknowledges the support of the Polish National Science Centre (NCN) grant no.\ 2017/27/B/ST2/02755. T.~S. kindly acknowledges the support of the Polish National Science Center (NCN) Grants No.\,2019/32/C/ST2/00202 and 2021/43/D/ST2/03375. 
S.~B. thanks for the hospitality at the Jagiellonian University where part of this work was done.
\end{acknowledgments}

\appendix

\section{Computation of the light-cone wave functions}
\label{sec:photonwf}

In order to compute the light-cone wave functions, the key element is a vertex contraction between spinors, e.~g.~$\bar{u}_h(k) \Gamma v_{\bar{h}}(k')$ ($\Gamma$ is some general Dirac vertex). For the spinors we use the LB basis, defined through
\be
\begin{split}
& u_h(k) = \frac{1}{2^{1/4}\sqrt{k^-}}\left(\sqrt{2}k^- + \gamma^0 m +  \boldsymbol{\alpha}_\perp \cdot \kp\right)\chi_h =  \frac{1}{2^{1/4}\sqrt{k^-}}\left(\sqrt{2}k^- + \gamma^0 m +  \slashed{\boldsymbol{k}}_\perp \gamma^0\right)\chi_h\,,\\
& v_h(k) = \frac{1}{2^{1/4}\sqrt{k^-}}\left(\sqrt{2}k^- - \gamma^0 m + \boldsymbol{\alpha}_\perp \cdot \kp\right)\chi_{-h} =\frac{1}{2^{1/4}\sqrt{k^-}}\left(\sqrt{2}k^- - \gamma^0 m + \slashed{\boldsymbol{k}}_\perp \gamma^0\right)\chi_{-h}\,,
\end{split}
\label{eq:spinor}
\ee
where $\alpha^i = \gamma^0 \gamma^i$. In accordance with the conventions used in this work, we have switched plus and minus light-cone coordinates in the above expression as compared to the original LB convention \cite{Lepage:1980fj}. Thus the spinors $\chi_h$ are eigenstates of $\gamma^+ \gamma^-$, namely
\be
\frac{1}{2}\gamma^+ \gamma^- \chi_h = \chi_h \,, \qquad \frac{1}{2}\gamma^- \gamma^+ \chi_h = 0\,.
\ee
Explicitly, we have $\chi_{+} = (-1,0,1,0)/\sqrt{2}$, $\chi_{-} = (0,1,0,1)/\sqrt{2}$. The spinors \eqref{eq:spinor} can now be written as
\be
\begin{split}
& u_h(k) = \frac{1}{2^{1/4}\sqrt{k^-}}(\slashed{k} + m)\gamma^0\chi_h\,,\\
& v_h(k) = \frac{1}{2^{1/4}\sqrt{k^-}}(\slashed{k} - m)\gamma^0\chi_{-h}\,.
\end{split}
\label{eq:blp}
\ee
By defining a projection matrix
\be
\chi_{h\bar{h}} \equiv \gamma^0\chi_{-\bar{h}}\bar{\chi}_{h}\gamma^0\,,
\label{eq:proj}
\ee
one obtains
\be
\bar{u}_h(k) \Gamma v_{\bar{h}}(k') = \frac{1}{\sqrt{2 k^- k'^-}}{\rm tr}\left[(\slashed{k}+ m) \Gamma (\slashed{k}' - m) \chi_{h\bar{h}}\right]\,,
\ee
and so the computation of the light-cone wave function comes down to the computation of the above Dirac trace. We use this method to compute all the wave functions considered in this work. Using \eqref{eq:proj}, the explicit forms of the projection matrices are found to be
\be
\begin{split}
&\chi_{\pm \pm} = \pm\frac{1}{2\sqrt{2}}\gamma^- (\gamma^1  \mp \rmi \gamma^2)\,,\\
&\chi_{\pm\mp} = \frac{1}{2\sqrt{2}}\gamma^-(1\mp\gamma_5)\,.
\end{split}
\ee

\section{Derivation of the amplitudes for axial and tensor quarkonia}
\label{appendix_overlaps_axial_tensor}

In this appendix we present the steps of the derivation of the amplitudes from Sec.~\ref{sec:FinalAmplitudes} for the axial vector and tensor quarkonia.

In the case of axial quarkonia we plug the second line of \eqref{eq:mesvert} into the traces in \eqref{eq:dirtr} to obtain
\be
\begin{split}
& A_{\lambda = 0, \lambdab = 0}(\lp,\lonp,z) = 0\,,\\
& A_{\lambda = 0,\lambdab = \pm}(\lp,\lonp,z) = 4  Q z\bar{z} (\epsp^{\bar{\lambda}*} \times \lonp)\,,\\
& A_{\lambda = \pm, \lambdab = 0}(\lp,\lonp,z) = \frac{4}{M_\calA}\left(m_c^2 (\epsp^\lambda \times \lonp)
 + \lonp^2(\epsp^\lambda \times \lp) \right)\,,\\
& A_{\lambda = \pm, \bar{\lambda} = \pm}(\lp,\lonp,z) = 2(z-\bar{z})\left((\epsp^\lambda\cdot\epsp^{\bar{\lambda}*})(\lp\times \lonp) - (\lp\cdot \lonp + m_c^2)(\epsp^\lambda\times \epsp^{\bar{\lambda}*})\right)\,,
\end{split}
\label{eq:diff}
\ee
The last line is understood to contain four different combinations of photon and quarkonia transverse polarizations. We used the following notations for the 2D cross products: $\xp \times \yp \equiv x^1 y^2 - x^2 y^1 = - \epsilon^{+-\xp\yp}$. The non-zero reduced amplitudes are 
\be
\begin{split}
&\calA_{\lambda = 0, \bar{\lambda} = \pm}(\rp,\delp) =  e q_c \rme^{-\rmi \bar{\lambda} \phi_r}\int_z \rme^{-\rmi \dlp\cdot \rp} \calA_{LT}(r_\perp)\,,\\
&\calA_{\lambda = \pm, \lambdab = 0}(\rp,\delp) = e q_c \rme^{\rmi \lambda\phi_r}\int_z \rme^{-\rmi \dlp\cdot \rp}\calA_{TL}(r_\perp)\,,\\
&\calA_{\lambda = \pm, \bar{\lambda} = \pm}(\rp,\delp) = eq_c\lambda\delta_{\lambda\lambdab}\int_z \rme^{-\rmi \dlp\cdot \rp}\calA_{TT}(r_\perp)\,,\\
\end{split}  
\label{eq:reduced_amplitudes}
\ee
where $\calA_A$ are given in Eqs.~(\ref{eq:Amps}). Notice that the first term from the fourth line of \eqref{eq:diff} is proportional to $\rp \times \rp = 0$ in coordinate space and so it 
does not contribute to $\calA_{TT}(r_\perp)$. In the last line of Eq.~\eqref{eq:reduced_amplitudes}, out of four possible combinations of 
photon and axial quarkonia transverse polarizations, only the two 
{polarization preserving transitions} survive since
this amplitude is proportional to the 2D cross product, $\epsp^\lambda\times \epsp^{\bar{\lambda}*} = 
(-\rmi/2)(\lambda+\bar{\lambda})= -\rmi\lambda\delta_{\lambda\bar{\lambda}}$ of the polarization vectors of the
incoming photon and outgoing axial vector quarkonia. {Accordingly, the two polarization flipping transitions vanish, unlike in case of vector quarkonia.}

Computing now the $\phi_r$ integral we separate the helicity dependence and obtain the amplitudes
\be
\begin{split}
& \left\langle \calM_{\lambda = 0,\lambdab = \pm 1}\right\rangle \equiv q^- \wcalM_{0,\lambdab = \pm 1} \equiv q^- \rme^{-\rmi\lambdab \phi_\Delta}\wcalM_{LT}\,,\\
& \left\langle \calM_{\lambda = \pm 1, \lambda = 0}\right\rangle \equiv q^- \wcalM_{\lambda = \pm 1, 0} \equiv q^- \rme^{\rmi \lambda \phi_\Delta}\wcalM_{TL}\,,\\
&\left\langle \calM_{\lambda = \pm 1,\lambdab = \pm 1}\right\rangle \equiv q^- \wcalM_{\lambda = \pm 1,\lambdab = \pm 1} \equiv q^- \lambda\delta_{\lambda\lambdab} \wcalM_{TT}\,,
\end{split}
\label{eq:Ms}
\ee
where $\wcalM_B$ are given in Eqs.~(\ref{eq:Mpol}).

For the tensor quarkonia, we proceed along similar steps as above, only this time there is a total of 15 overlaps that need to be computed. Starting from the traces \eqref{eq:dirtr}, we plug in the third line of \eqref{eq:mesvert} and find 
\be
\begin{split}
 A_{\lambda = 0,\lambdab = \pm 2}(\lp,\lonp,z) &= - 8 Q z\bar{z}(z-\bar{z})(\epsp^{\pm *}\cdot \lonp)^2\,,\\ 
 A_{\lambda = \pm 1,\lambdab = \pm 2}(\lp,\lonp ,z) &= -4 (\epsp^{\pm*}\cdot\lonp)\big[(\epsp^\lambda\cdot \epsp^{\pm *})(\lp\cdot\lonp + m_c^2)\\
 & + (z-\bar{z})^2 (\epsp^\lambda\cdot \lp) (\epsp^{\pm *}\cdot \lonp) - (\epsp^\lambda\cdot \lonp)(\epsp^{\pm *}\cdot \lp)\big]\,,\\
 A_{\lambda = 0,\lambdab = \pm 1}(\lp,\lonp ,z) & = 2\sqrt{2} Q M_\calT  z\bar{z}(3-4(z^2 +\bar{z}^2))(\epsp^{\lambdab*}\cdot \lonp)\,,\\
 A_{\lambda = \pm 1,\lambdab = \pm 1}(\lp,\lonp ,z) & = -\sqrt{2} M_\calT (z-\bar{z}) \big[(\epsp^\lambda\cdot\epsp^{\lambdab *})(\lp\cdot\lonp + m_c^2)\\
 & - (3 - 4(z^2 + \bar{z}^2))(\epsp^\lambda \cdot \lp)(\epsp^{\lambdab *}\cdot\lonp) - (\epsp^\lambda\cdot\lonp)(\epsp^{\lambdab *}\cdot \lp)\big]\,,\\
A_{\lambda = 0,\lambdab = 0}(\lp,\lonp ,z) & = \frac{8\sqrt{2}Q}{\sqrt{3}}z\bar{z}(z-\bar{z})\left(\frac{3}{2}\lonp^2 + m_c^2\right)\,,\\
  A_{\lambda = \pm 1,\lambdab = 0}(\lp,\lonp ,z) & = \frac{ 2\sqrt{2}}{\sqrt{3}}\Bigg[2(z-\bar{z})^2(\epsp^\lambda \cdot \lp)\left( \frac{3}{2}\lonp^2 + m_c^2\right) + (\epsp^\lambda \cdot \lonp)  m_c^2\Bigg]\,.
\end{split}
\label{eq:tenstr}
\ee
The results for $\lambdab = 0$ have been simplified using
\be
(\epsp^{+*} \cdot\up)(\epsp^{-*} \cdot\vp) + (\epsp^{-*} \cdot\up) (\epsp^{+*} \cdot\vp) = - \up \cdot\vp\,,\\
\ee
which holds for general 2D vectors $\up$ and $\vp$. Inserting \eqref{eq:tenstr} in \eqref{eq:redA} we obtain the reduced amplitudes
\be
\begin{split}
&\calA_{\lambda = 0, \lambdab = \pm 2}(\rp,\delp)  = e q_c \rme^{-\rmi \lambdab \phi_r}\int_z \rme^{-\rmi \dlp\cdot \rp} \calA_{LT2}(r_\perp)\,,\\
&\calA_{\lambda = \pm 1, \lambdab = \pm 2}(\rp,\delp) = e q_c \lambda \rme^{\rmi (\lambda - \lambdab)\phi_r} \int_z \rme^{-\rmi \dlp\cdot \rp}\left( \delta_{\lambda,\lambdab/2}\calA_{TT2p}(r_\perp) + \delta_{\lambda,-\lambdab/2} \calA_{TT2f}(r_\perp)\right)\,,\\
&\calA_{\lambda = 0, \lambdab = \pm 1}(\rp,\delp) = e q_c \lambdab \rme^{-\rmi\lambdab \phi_r}\int_z \rme^{-\rmi \dlp\cdot \rp} \calA_{LT}(r_\perp)\,,\\
&\calA_{\lambda = \pm 1, \lambdab = \pm 1}(\rp,\delp) = e q_c \rme^{\rmi (\lambda - \lambdab)\phi_r}\int_z \rme^{-\rmi \dlp\cdot \rp} \left( \delta_{\lambda\lambdab}\calA_{TTp}(r_\perp) - \delta_{\lambda,-\lambdab}\calA_{TTf}(r_\perp)\right)\,,\\
&\calA_{\lambda = 0, \lambdab = 0}(\rp,\delp) = e q_c \int_z \rme^{-\rmi \dlp\cdot \rp} \calA_{LL}(r_\perp)\,,\\
&\calA_{\lambda = \pm 1, \lambdab = 0}(\rp,\delp) = e q_c {\lambda\rme^{\rmi \lambda \phi_r}} \int_z \rme^{-\rmi \dlp\cdot \rp} \calA_{TL}(r_\perp)\, .\\
\end{split}
\label{eq:redcT}
\ee 

From \eqref{eq:redcT} we see that for the transverse polarizations of the photon and the tensor quarkonia both the polarization preserving ($\lambda \to \lambdab/2$, for the case $\lambdab = \pm 2$ and $\lambda \to \lambdab$ for the case $\lambdab = \pm 1$) and polarization flipped ($\lambda \to - \lambdab/2$, $\lambda \to -\lambdab$) transitions are allowed, which explains the notation ($p =$ preserving, $f$ = flipped). 

In the final step we compute the amplitudes
\be
\begin{split}
&\left\langle \calM_{\lambda = 0,\bar{\lambda} = \pm 2}\right\rangle \equiv q^- \wcalM_{\lambda = 0, \bar{\lambda} = \pm 2} \equiv q^- \rme^{-\rmi \bar{\lambda} \phi_\Delta} \wcalM_{LT2}\,,\\
&\left\langle \calM_{\lambda = \pm 1,\bar{\lambda} = \pm 2}\right\rangle \equiv q^- \wcalM_{\lambda = \pm 1, \bar{\lambda} = \pm 2} \equiv q^- \lambda\rme^{-\rmi\lambda \phi_\Delta}\delta_{\lambda,\lambdab/2} \wcalM_{TT2p} + q^- \lambda\rme^{3\rmi\lambda \phi_\Delta}\delta_{\lambda,-\lambdab/2}\wcalM_{TT2f}\,,\\
&\left\langle \calM_{\lambda = 0,\bar{\lambda} = \pm 1}\right\rangle \equiv q^- \wcalM_{\lambda = 0, \bar{\lambda} = \pm 1} \equiv q^- \rme^{-\rmi\lambdab \phi_\Delta}\wcalM_{LT}\,,\\
&\left\langle \calM_{\lambda = \pm 1,\bar{\lambda} = \pm 1}\right\rangle \equiv q^-\wcalM_{\lambda = \pm 1, \bar{\lambda} = \pm 1} \equiv q^- \delta_{\lambda \lambdab}\wcalM_{TTp} + q^- \delta_{\lambda,-\lambdab}\rme^{2\rmi\lambda\phi_\Delta } \wcalM_{TTf}\,,\\
&\left\langle \calM_{\lambda = 0,\bar{\lambda} = 0}\right\rangle \equiv q^-\wcalM_{\lambda = 0, \bar{\lambda} = 0} \equiv q^-\wcalM_{LL}\,,\\
&\left\langle \calM_{\lambda = \pm 1,\bar{\lambda} =  0}\right\rangle \equiv q^- \wcalM_{\lambda = \pm 1, \bar{\lambda} = 0} \equiv q^- (\delta_{\lambda 1} - \delta_{\lambda,-1})\rme^{\rmi \lambda\phi_\Delta} \wcalM_{TL}\,,
\end{split}
\label{eq:AmpsTens}
\ee
where the explicit helicity dependence was also factored out. The scalar functions $\wcalM_B$ are given by Eqs.~(\ref{eq:MpolTens}).

\section{The Primakoff contribution in specific kinematic limits}
\label{primakoff_limits}

\subsection{The proof of formula (\ref{eq:m-general})}
\label{sec:scaltens}

We begin by considering the amplitude for the decay process $\calH(\lambdab,\Delta)\to\gamma(\lambda,q)\gamma(\lambda',\ell)$, where $\calH = \calS$, $\calT$. In what follows, it will be useful to separate out the exchanged photon polarization from the amplitude as
\be
\calM_{\lambda \lambda' \lambdab}(\calH\to\gamma\gamma) = \epsilon_\mu(\lambda',\ell)\calM^\mu_{\lambda\lambdab}(\calH\to\gamma\gamma)\,.
\label{eq:2gammadec}
\ee
Inserting \eqref{eq:2gammadec} into the standard formula for the decay width \cite{ParticleDataGroup:2022pth} we get 
\be
\begin{split}
\Gamma(\calH \to \gamma\gamma) & = \frac{1}{2J+1}\frac{1}{32\pi  M_{\calH} }\sum_{\lambdab=-J} ^J \,\sum_{\lambda \lambda'} |\calM_{\lambda \lambda' \lambdab}(\calH\to\gamma\gamma)|^2\,,\\
& = -\frac{1}{2 J + 1}\frac{1}{32 \pi M_\calH}\sum_{\lambdab = - J}^J \sum_\lambda g_{ii'}\calM^i_{\lambda\lambdab}(\calH \to \gamma\gamma)\calM^{i'*}_{\lambda\lambdab}(\calH \to \gamma\gamma)\,.
\label{eq:Gamma-m}
\end{split}
\ee

Next, we consider the amplitude for exclusive photoproduction $\calM_{\lambda\lambdab} (\gamma p \to \calH p)$, see \eqref{eq:amppauli}. 
The cross section reads~\cite{ParticleDataGroup:2022pth}
\be
\frac{\rmd\sigma_{\calH}(\gamma p \to \calH p)}{\rmd |t|} = \frac{1}{2} \sum_{\lambdab=-J} ^J\sum_{\lambda}\int\frac{\rmd \phi_\ell}{2\pi} 
\frac{1}{16\pi W^4}
\left|\calM_{\lambda\lambdab} (\gamma p \to \calH p)\right|^2 \, .
\label{eq:xsec}
\ee
where $W^2=(q+P)^2 \simeq 2 q\cdot P \gg M_{\calH} ^2, \, P^2, \,|\ell^2|$ in the high energy limit\footnote{The amplitudes $\calM_{\lambda\bar{\lambda}}$ in this section and the amplitudes $\wcalM_{\lambda\lambdab}$ of
Eqs.~\eqref{eq:Mscal}, \eqref{eq:Ms} and \eqref{eq:AmpsTens} are related through $\wcalM_{\lambda\lambdab} = \calM_{\lambda\lambdab}/W^2$, compare \eqref{eq:xsec} with \eqref{eq:csmain}.}. Moreover $\ell \approx (x_\calP P^+,0, \ellp)$ with $\phi_\ell$ the azimuthal angle of $\ellp$. We also have $x_\calP \approx M_{\calH} ^2 /W^2$ (see Eq.~\eqref{eq:pomx}), and $\ell^2 = t \approx -\ellp^2$.

Taking the high energy limit leads to \eqref{eq:amppauli2}. Here we pick up the leading $F_1$ term when $W^2 \gg |t|$ 
\be
\calM_{\lambda\lambdab} (\gamma p \to \calH p) = P_{\mu} \, \calM_{\lambda\lambdab}^\mu(\gamma\gamma \to \calH) \frac{e F_1(\ell_\perp)}{\ell^2} \, \frac{1}{q\cdot P} \bar u(P') \slashed{q} u(P)\,.
\label{eq:amp7}
\ee
To make the connection \eqref{eq:m-general} we rewrite \eqref{eq:amp7} using QED gauge invariance $\ell_\mu \calM_{\lambda\lambdab}^\mu(\gamma\gamma\to\calH) = 0$ in the high-energy limit (the Collins-Ellis trick \cite{Collins:1991ty})
\be
P_{\mu} \calM_{\lambda\lambdab}^\mu(\gamma\gamma \to \calH) \approx  -\frac{1}{x_\calP}\ell_i \calM_{\lambda\lambdab}^i(\gamma\gamma \to \calH)\,,
\ee
so that
\be
\calM_{\lambda\lambdab} (\gamma p \to \calH p) = - \frac{ 2e F_1(\ell_\perp)}{t} \frac{W^2}{M^2 _{\calH}}
\ell_i \calM_{\lambda\lambdab}^i(\gamma\gamma \to \calH).
\label{eq:Mge2me}
\ee
Finally, the result in \eqref{eq:xsec} is isotropic in $\phi_\ell$ after the polarization sum. We can perform the angular average leading to
\be
\begin{split}
\frac{1}{2}\sum_{\lambdab = - J}^J\sum_\lambda \int \frac{\rmd \phi_\ell}{2\pi} \left|\ell_i \calM_{\lambda\lambdab}^i(\gamma\gamma \to \calH)\right|^2 & = -\frac{1}{4}\ellp^2 \sum_{\lambdab = - J}^J \sum_\lambda g_{ii'}\calM_{\lambda\lambdab}^i(\gamma\gamma \to \calH)\calM_{\lambda\lambdab}^{i' *}(\gamma\gamma \to \calH)\\
& =  (2 J + 1)8\pi M_\calH \ellp^2 \Gamma(\calH \to \gamma\gamma)\,,
\end{split}
\label{eq:relation}
\ee
where in the last line we used \eqref{eq:Gamma-m}.
Inserting \eqref{eq:relation} into \eqref{eq:xsec} via \eqref{eq:Mge2me} gives Eq.~\eqref{eq:m-general}, which is the desired result. In the derivation we assumed scattering off a proton target, but the formula is valid for any charged particle.

\subsection{The $|t|\to 0$ limit for axial vector quarkonia and its connection to the Landau-Yang theorem}
\label{sec:axial}

Thanks to the Collins-Ellis trick, the $\gamma^*\gamma^*\to \calH$ amplitude for scalar and tensor quarkonia has a finite $O(\Delta_\perp)$ contribution. The Collins-Ellis trick is a rather general and robust procedure in the high-energy limit and so from this perspective the case of axial quarkonia seems special in that the $O(\Delta_\perp)$ contribution vanishes, with the amplitude scaling as $O(\Delta_\perp^2)$, check \eqref{eq:MAt0}. In the following we demonstrate that $O(\Delta_\perp^2)$ scaling follows from a general argument which does not rely on a particular choice of axial quarkonium wave function.

We take Eq.~\eqref{eq:2gamma} as a starting point and further factor out the polarization vectors from the helicity amplitude as
\be
n_\nu\calM^\nu_{\lambda \lambdab}(\gamma^*\gamma^*\to\calA) = \epsilon_\mu(\lambda,q)n_\nu E_\rho(\lambdab,\Delta_0)\calM^{\mu\nu\rho}(\gamma^*\gamma^*\to\calA)\,.
\label{eq:mprim}
\ee
The covariant amplitude $\calM^{\mu\nu\rho}(\gamma^*\gamma^*\to\calA)$ depends only on the vectors $q^\mu$ and $\ell^\nu$. Because $\calA$ is an axial vector quarkonium, $\calM^{\mu\nu\rho}(\gamma^*\gamma^*\to\calA)$ has to be proportional to the $\epsilon$-tensor. In general the covariant decomposition is as follows 
\be
\begin{split}
\calM^{\mu\nu\rho}(\gamma^*\gamma^*\to\calA) & = \left(q^\rho - \ell^\rho + \frac{-q^2 + \ell^2}{(q + \ell)^2}(q^\rho + \ell^\rho)\right) \epsilon^{\mu\nu q\ell} M_\calA  F_{TT}(q^2,\ell^2)\\
& + \left(\ell^\mu - \frac{q\cdot\ell}{q^2} q^\mu\right)\epsilon^{\nu\rho q \ell} \sqrt{-q^2} F_{LT}(q^2,\ell^2) + \left(q^\nu - \frac{q\cdot\ell}{\ell^2} \ell^\nu\right)\epsilon^{\mu\rho q \ell} \sqrt{-\ell^2} F_{TL}(q^2,\ell^2)\,,
\end{split}
\label{eq:Mcov}
\ee
see  e.~g.~\cite{Babiarz:2022xxm} and references therein. The decomposition \eqref{eq:Mcov} is QED gauge invariant, that is: $q_\mu \calM^{\mu\nu\rho}(\gamma^*\gamma^*\to \calA) = 0$ and $\ell_\nu \calM^{\mu\nu\rho}(\gamma^*\gamma^*\to \calA) = 0$, and we also have $(q^\rho + l^\rho) \calM_{\mu\nu\rho}(\gamma^*\gamma^*\to\calA) = 0$. Here the notation $F_{TT}$, $F_{TL}$ and $F_{LT}$ refers to the $\gamma^*\gamma^*$ polarizations in the center of mass frame. 
Only the form-factor $F_{TT}$ is constrained by the LY theorem: $F_{TT}(0,0) = 0$ \cite{Landau:1948kw,Yang:1950rg}, see also ref.~\cite{Babiarz:2022xxm}. Due to the Bose symmetry of the full amplitude \eqref{eq:Mcov} we must have $F_{TT}(q^2,\ell^2) = - F_{TT}(\ell^2, q^2)$ and so $F_{TT}(q^2,\ell^2) \propto q^2 - \ell^2$ for small $q^2$ and $\ell^2$. On the other hand, the form-factors $F_{LT}$ and $F_{TL}$ scale as $F_{LT}(q^2,\ell^2) \propto \sqrt{-q^2}$, $F_{TL}(q^2,\ell^2) \propto \sqrt{-\ell^2}$ in order to avoid kinematic singularities \cite{TPCTwoGamma:1988izb,Babiarz:2022xxm}.

Contracting \eqref{eq:Mcov} with the longitudinal polarization vector $E^\rho(0,\Delta_0)$, only the piece proportional to the gauge vector $n^\rho$ from \eqref{eq:Elong} survives. Further contracting with the transverse photon polarization $\epsilon^\mu(\lambda = \pm 1,q)$ and the gauge vector $n^\nu$ we obtain
\be
n_\nu \calM_{\lambda = \pm 1, \lambdab = 0}^\nu(\gamma^*\gamma^* \to \calA)  = -\frac{M_0}{\Delta_0\cdot n}(q\cdot n)\epsilon^{\epsilon(\lambda,q) n q \ell}\left[\left(1 - \frac{q^2 - \ell^2}{\Delta_0^2} \right) M_{\calA} F_{TT}(q^2,\ell^2) + \sqrt{-\ell^2} F_{TL}(q^2,\ell^2)\right]\,,
\label{eq:Mlongcov}
\ee
where we have used $\ell \cdot n = 0$. The form-factor $F_{LT}(q^2,\ell^2)$ decouples after contracting $\epsilon^{\nu\rho q l}$ with $n_\nu E_\rho(0,\Delta_0)$. Taking into account the scalings of the form-factors the square bracket in \eqref{eq:Mlongcov} is $O(\Delta_\perp^2)$ while the $\epsilon$-tensor is $O(\Delta_\perp)$ leading to the overall $O(\Delta_\perp^3)$ scaling. 

For the transverse polarization we find
\be
\begin{split}
n_\nu \calM_{\lambda = \pm 1, \lambdab = \pm 1}^\nu(\gamma^*\gamma^* \to \calA) & = (q-l)\cdot E(\lambdab,\Delta_0) \epsilon^{\epsilon(\lambda,q) n q \ell} M_{\calA} F_{TT}(q^2,\ell^2) + (\epsilon(\lambda,q)\cdot \ell) \epsilon^{n E(\lambdab,\Delta_0) q \ell} \sqrt{-q^2} F_{LT}(q,\ell^2)\\
& + (q\cdot n) \epsilon^{\epsilon(\lambda,q)E(\lambdab,\Delta_0)q \ell} \sqrt{-\ell^2}F_{TL}(q^2,\ell^2)\,. 
\end{split}
\label{eq:Mtranscov}
\ee
The first term in \eqref{eq:Mtranscov} is $O(\Delta_\perp^3)$, the second term decouples in the $q^2 \to 0 $ limit and the third term is $O(\Delta_\perp^2)$ which is the leading contribution in the $\Delta_\perp \to 0$ limit.

\subsection{The NRQCD limit of the Primakoff cross sections}

In this subsection we consider the non-relativisitic QCD (NRQCD) limit of the Primakoff process by taking
the heavy quark limit. We also take $Q^2 \to 0$ and focus on transverse photon - proton
scattering.

In the NRQCD limit all the momenta in the amplitude are considered to be much smaller than the heavy quark mass. Since the expressions obtained in the previous Secs.~\ref{sec:scaltens} and \ref{sec:axial} already correspond to the $|t| \to 0$ limit, we take those as starting point and further expand to second order
in $\xi = z - 1/2$ and $\lp$.

For scalar quarkonia, using \eqref{eq:MSt0} we obtain
\be
\wcalM_T \to 8\sqrt{2}\pi \rmi q_c^2 \alpha N_c \frac{e F_1(\Delta_\perp)}{\Delta_\perp} \frac{1}{m_c^3}\int_z\int_{\lp} (\lp^2 + 4\xi^2 m_c^2)\frac{\phi_\calS(\lp,z)}{z\bar{z}}\,.
\label{eq:mnrqcd}
\ee
In the next step we replace the LCWF $\phi_\calS(\lp,z)$ by the non-relativistic
radial wave function $u(l)$. Here $\lth = (\lp,l^3)$ and $l$ is its modulus, with $l^3 = 2\xi m_c$ as appropriate for the NRQCD limit. The correspondence is given by~\cite{Babiarz:2020jkh}
\be
\int_z \int_{\lp} \frac{\phi_\calS(\lp,z)}{z\bar{z}} \to 
\frac{1}{16\pi^2 \sqrt{m_c}} \frac{1}{\sqrt{N_c}} \int \rmd^3 \lth \frac{u(l)}{\lth^2} \,.
\label{eq:corresp}
\ee
After performing the angular integrations in Eq.~\eqref{eq:mnrqcd} we relate
$u(l)$ to the derivative of the radial wave function at the origin, as in Eq.~(3.18) of Ref.~\cite{Babiarz:2020jkh}:
\be
\int_0^{\infty} \rmd l \, l^2 u(l) = 3\sqrt{\frac{\pi}{2}} R'(0)\,.
\label{eq:dR0}
\ee
This leads to
\be
\wcalM_T \to  12\sqrt{\pi} \rmi q_c^2 \alpha \sqrt{N_c} \frac{e F_1(\Delta_\perp)}{\Delta_\perp} \frac{1}{m_c^{ 7/2}} R'(0)\,.
\ee

The cross section is obtained from standard formulas, see Eq.~\eqref{eq:csS}. 
We find
\be
\frac{\rmd\sigma(\gamma p \to \calS p)}{\rmd |t|} \to \frac{9\pi q_c^4 \alpha^3 N_c |R'(0)|^2 F_1^2(0)}{m_c^7 |t|}\,.
\label{eq:csnrS}
\ee
The result agrees with Eq.~(8b) of Jia {\it et al.}~\cite{Jia:2022oyl}.

For axial vector quarkonia there is an additional factor of $\sqrt{3/2}$ on the r.h.s.\ of
Eq.~(\ref{eq:corresp}). Recalling that the leading contribution in the $|t| \to 0$ limit comes from the amplitude $\wcalM_{TT}$ of \eqref{eq:MAt0}, and using \eqref{eq:csA}, we find that in the NRQCD limit
\be
\frac{\rmd\sigma(\gamma p \to \calA p)}{\rmd |t|} \to \frac{3\pi q_c^4 \alpha^3 N_c |R'(0)|^2 F_1^2(0)}{m_c^9}\, ,
\ee
which also agrees with Eq.~(8c) in Ref.~\cite{Jia:2022oyl}.

For the tensor quarkonia with $\bar\lambda= \pm 1, \pm 2$, the r.h.s.\ of Eq.~\eqref{eq:corresp}
is multiplied by a factor of $\sqrt{3}/M_\calT$. Using \eqref{eq:MTt0} and \eqref{eq:csT} we find that the leading contribution comes from the amplitude $\wcalM_{TT2,p}$ so that
\be
\frac{\rmd\sigma(\gamma p \to \calT p)}{\rmd |t|} \to \frac{12\pi q_c^4 \alpha^3 N_c |R'(0)|^2 F_1^2(0)}{m_c^7 |t|}\, .
\ee
This also agrees with Eq.~(8d) in Ref.~\cite{Jia:2022oyl}.

\section{The relative sign of the eikonal photon and Odderon exchange amplitudes}
\label{sec:Odderon-sign}

In this appendix we explain in detail our conventions for covariant
derivatives, field equations, and Wilson lines. Consistent conventions
are important for obtaining the correct relative sign of the Primakoff
and Odderon amplitudes.

We write the covariant derivative in the fundamental representation as
\be
(D^\mu)_{ij} = \delta_{ij}\left(\partial^\mu + \rmi e q_c A^\mu\right)
+  \rmi g A^{\mu a} (t^a)_{ij}~.
\ee
Here, $i, j = 1,\dots, N_c$ are fundamental and $a=1,\dots, (N_c^2-1)$
is an adjoint color index; $A^\mu, A^{\mu a}$ represent the
electromagnetic and color fields, respectively, and $t^a$ are the
traceless generators of color-SU$(N_c)$, normalized as $\mathrm{tr}(t^a
t^b) = \frac{1}{2}\delta^{ab}$.  $q_c$ denotes the fractional
electromagnetic charge of the fermion field on which this covariant
derivative acts, $q_c=+2/3$ for $c$ quarks.

From the Dirac equation for a massless fermion field, $\rmi \gamma_\mu (D^\mu)_{ij} \psi_j = 0$, in a background with eikonal (``shockwave'') $A^\mu$ and $A^{\mu a}$ fields (where $A^+$ is the only
non-vanishing field component and independent of $x^+$), $\psi_j$ is proportional to the Wilson line
\be
V(\xp) = {\cal P}\, \exp\left\{
  - \rmi  \int_{-\infty}^{+\infty} \rmd x^- \left[e q_c A^{+}(x^-,\xp) + g A^{+ a}(x^-,\xp) t^a\right]
  \right\}~.
\label{eq:Wilson_Line}
\ee
The Yang-Mills and Maxwell equations for the shockwave fields in covariant gauge are
\begin{eqnarray}
\partial_\mu F^{\mu\nu}_a = J^\nu_a &\to&
-  \boldsymbol{\nabla}_\perp^2 A^{+}_a = J^+_a~, \\
\partial_\mu F^{\mu\nu} = J^\nu &\to&
- \boldsymbol{\nabla}_\perp^2 A^{+} = J^+~,
\label{eq:fields}
\end{eqnarray}
where $J^{+ a}$ and $J^{+}$
are the plus components of the
color and electromagnetic currents, respectively. Note that here
$A^+$ denotes the electromagnetic field sourced by the charges in the proton,
from which the projectile $c\bar{c}$ dipole scatters.

We now define the $S$-matrix for eikonal scattering of the color singlet
dipole, averaged over the colors of the $c, \bar c$ quarks
\be \label{eq:Sdipole-V-Vdagger}
S(\xp, \yp) = \frac{1}{N_c}\, {\rm tr}\left\langle
V(\xp) V^\dagger(\yp)\right\rangle\,.
\ee
$\langle\cdots\rangle$ denotes the matrix element of the respective
operator between proton states $|P^+,\boldsymbol{P}_\perp\rangle$ and
$\langle P^+ + \Delta^+,\boldsymbol{P}_\perp+\boldsymbol{\Delta}_\perp|$, stripped of the
$\delta$-functions representing conservation of light-cone and
transverse momentum.

We can now define the amplitudes for single photon or three gluon
exchange as follows. Setting the QCD coupling $g=0$ and expanding
the Wilson lines to linear order in $A^+$ we have
\be
V(\xp) = 1-\rmi e q_c \int_{-\infty}^{+\infty} \mathrm{d}x^- A^{+}(x^-,\xp)
  \equiv 1- \rmi e q_c \alpha(\xp)\,.
\ee
Then,
\be
\Omega(\xp, \yp) \equiv - \frac{1}{2\rmi} \left\langle V(\xp)V^\dagger(\yp)
  - V(\yp)V^\dagger(\xp)\right\rangle =
e q_c \left\langle\alpha(\xp) - \alpha(\yp)\right\rangle\,.
\ee
corresponds to the scattering amplitude for single photon exchange.
With the field equation~(\ref{eq:fields}) we can write
$\langle\alpha(\xp)\rangle = - \left\langle \frac{1}{\boldsymbol{\nabla}^2_\perp}\rho(\xp) \right\rangle$
where the integrated electric charge density is $\rho(\xp) = \int_{-\infty}^{+\infty}\rmd x^- J^+(x^-,\xp)$.
Performing a Fourier transform from the transverse coordinate to the transverse
momentum space, the matrix element of this operator is simply the
Dirac electromagnetic form factor of the proton, $e F_1(q_\perp^2) =
\langle \rho(\qp) \rangle$. 
This leads to
\be
\Omega(\rp, \delp) = 8\pi\rmi \, \alpha q_c \frac{F_1(\Delta_\perp^2)}{\delp^2}
\, \sin \left(\frac{\rp \cdot\delp}{2}\right)~.
\ee
Here we introduced $\rp = \xp - \yp$. This is the expression
written in Eq.~(\ref{eq:photonXchangeAmpli}) of the main text.
In impact parameter space,
\be
\Omega(\rp, \bp) = 8\pi \alpha q_c \int\limits_{\delp}\sin(\bp \cdot \delp)\, 
\frac{- F_1(\Delta_\perp^2)}{\delp^2}
\, \sin\left(\frac{\rp \cdot \delp}{2}\right)~.
\label{eq:Omega_r_b}
\ee

\begin{figure}[htb]
  \begin{center}
  \includegraphics[scale = 0.8]{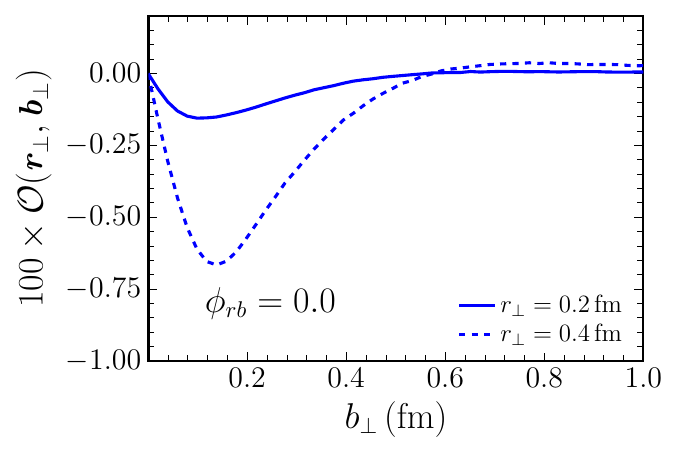}
  \end{center}
  \caption{The Odderon amplitude $\calO(\rp, \bp)$ obtained from a light-front quark model~\cite{Schlumpf:1992vq,Brodsky:1994fz},
  when the relative angle between $\rp$ and $\bp$ $\phi_{rb} = 0$.
  At small $b_\perp$ the sign is negative.
  }
  \label{fig:odderBS}
\end{figure}
We now proceed to the three gluon exchange amplitude by setting $q_c=0$ in the Wilson
line~(\ref{eq:Wilson_Line}) followed by an expansion of \eqref{eq:odd}, namely
\be  \label{eq:O-V-Vdagger}
\calO(\xp, \yp) = - \frac{1}{2\rmi N_c} \, \mathrm{tr}\, \left\langle V(\xp)V^\dagger(\yp)
  - V(\yp)V^\dagger(\xp)\right\rangle\,,
\ee
to third order in $A^{+a}$:
\be
\begin{split}
    \calO(\rp, \bp) & = \frac{g^3}{4N_c} d^{abc} \int_{\qonp \qtwp \qthp}
    \frac{\langle\rho^a(\qonp)\rho^b(\qtwp)\rho^c(\qthp)\rangle}{\qonp^2 \qtwp^2 \qthp^2}
    \sin\left(\bp \cdot (\qonp + \qtwp + \qthp)\right)\\ 
    & \times \left[ 
    \sin\left(\frac{1}{2}\rp\cdot (\qonp - \qtwp - \qthp)\right) + 
    \frac{1}{3}\sin\left(\frac{1}{2}\rp \cdot (\qonp + \qtwp + \qthp)\right)\right]~.
\end{split}
\label{eq:Orb_G3}
\ee
For details see Eq.~(77) of Ref.~\cite{Dumitru:2018vpr}, to be multiplied by a factor of $\rmi$ (compare our Eq.~(\ref{eq:O-V-Vdagger}) to their
Eq.~(76)), and their Appendix~B.
Here $\rho^a(\xp) = \int_{-\infty}^{+\infty}\mathrm{d}x^- J^{+a}(x^-, \xp)$ denotes the color
charge density integrated along the eikonal path.
We now write the $C$-conjugation odd part of the color charge correlator as
\be
\left\langle\rho^a(\qonp)\rho^b(\qtwp)\rho^c(\qthp)\right\rangle = \frac{1}{4} d^{abc} g^3\, G_3^-(\qonp, \qtwp, \qthp)~.
\ee
In a three-quark model this correlator takes the form~\cite{Dumitru:2018vpr}
\begin{eqnarray}
G_3^-(\qonp, \qtwp, \qthp)
&=&
\int [\mathrm{d} x_i]  \int [\mathrm{d}^2 p_i] \, \psi(x_1,\ponp; x_2,\ptwp; x_3,\pthp)\nonumber\\
& &
\Bigl[ 
\psi^*(x_1,\ponp+(x_1-1)\qp; x_2,\ptwp+x_2\qp;
x_3,\pthp+x_3\qp) \nonumber\\
& & -
  \psi^*(x_1,\ponp-\qonp +x_1\qp; x_2,\ptwp-\qtwp-\qthp
  + x_2\qp; x_3, \pthp+x_3\qp) \nonumber\\
  & &
  -
 \psi^*(x_1,\ponp-\qonp-\qthp +x_1\qp; x_2, \ptwp-\qtwp
  + x_2\qp; x_3, \pthp+x_3\qp) \nonumber\\ & & -
 \psi^*(x_1,\ponp-\qonp-\qtwp +x_1 \qp;
  x_2,\ptwp-\qthp+ x_2\qp;
  x_3,\pthp+x_3\qp) \nonumber\\ & & +2\, 
\psi^*(x_1,\ponp-\qonp +x_1\qp; x_2,\ptwp-\qtwp +x_2 \qp;
  x_3,\pthp-\qthp + x_3\qp ) \Bigr]\,  \label{eq:rho^3correl}~.
\end{eqnarray}
Here, $\qp=\qonp+\qtwp+\qthp$ and
\be
\begin{split}
& [\mathrm{d} x_i] \equiv \rmd x_1 \rmd x_2 \rmd x_3 \delta(1-x_1 - x_2 - x_3)\,,\\
& [\mathrm{d}^2 p_i] \equiv \frac{1}{(16\pi^3)^2}\rmd^2 \ponp \rmd^2  \ptwp \rmd^2 \pthp \delta^{(2)}(\ponp + \ptwp + \pthp)\,,
\end{split}
\ee 
denote integrations over the quark light-cone momentum fractions and transverse momenta. $\psi(x_1,\ponp;x_2,\ptwp;x_3,\pthp)$ is the light-cone wave
function of the model proton.
This expression agrees with the $C$-odd three-gluon
exchange proton impact factor $E_{3;0}$ by Bartels and
Motyka~\cite{Bartels:2007aa}
up to a conventional factor of $(-\rmi)^3$.
The first term in Eq.~(\ref{eq:rho^3correl}) corresponds to the coupling of the three
exchanged gluons to the same quark in the proton, and it is equal to the Dirac form factor $F_1(\qp^2)$;
note the opposite sign as compared to the photon exchange amplitude~(\ref{eq:Omega_r_b}). This contribution
is dominant when all $|\boldsymbol q_{\perp i}|$ are much greater than the typical quark transverse momentum while
$|\qp|$ is on the order of that scale. For high momentum transfer $|t|=\qp^2$ on the other hand,
contributions due to gluon exchanges with two or three quarks in the proton become dominant.
Where this transition occurs is determined by the light-cone wave function $\psi$ which encodes
the structure of the proton, specifically
the single quark momentum distribution as well as multi-quark momentum correlations.

$G_3^-$ can also be expressed~\cite{Dumitru:2020fdh} in terms of two-gluon exchanges,
$G_2(\qonp, \qtwp)=
2\delta^{ab}\left<\rho^a(\qonp)\, \rho^b(\qtwp)\right>/g^2(N_c^2-1)$, where two of the three gluons are ``paired up'',
plus a genuine 3-body contribution which enforces the Ward identity
(vanishing of $G_3^-$) when either one of the transverse momenta vanishes, $\boldsymbol q_{\perp i} \to 0$:
\begin{eqnarray}
G_3^-(\qonp, \qtwp, \qthp) &=&
G_2(\qonp+ \qtwp, \qthp) 
+G_2(\qonp+ \qthp, \qtwp) 
+G_2(\qtwp+ \qthp, \qonp) \nonumber\\
& & -2\int [\mathrm{d} x_i] \int [\mathrm{d}^2 p_i] \left[
\psi^*(x_1,\ponp+(x_1-1)\qp; x_2,\ptwp+x_2\qp;
x_3,\pthp+x_3\qp) \right. \nonumber\\
& & ~~~~~~~~~ \left. -\psi^*(x_1,\ponp-\qonp +x_1\qp; x_2 ,\ptwp-\qtwp +x_2 \qp; x_3,
  \pthp-\qthp + x_3\qp ) \right] \times \nonumber\\
  \, & & ~~~~~~~~~~~~~~~~~~ \psi(x_1,\ponp; x_2, \ptwp; x_3 ,\pthp)~.
\end{eqnarray}
Once again, for small $|\qp|$ but large $|\boldsymbol q_{\perp i}|$ the fourth term on the right-hand-side of this equation becomes
$-2F_1(\qp^2)$.

A simple light-front quark model wave function
from the literature~\cite{Schlumpf:1992vq,Brodsky:1994fz} predicts $\calO(\rp, \bp)<0$ at
small impact parameters, see Fig.~\ref{fig:odderBS}. This corresponds to constructive
interference of photon and Odderon exchange amplitudes.
Neither the fixed order ${\cal O}(\alpha_S)$ correction~\cite{Dumitru:2022ooz} 
to the matrix element $\langle\rho^a(\qonp)\rho^b(\qtwp)\rho^c(\qthp)\rangle$ nor
small-$x$ resummation of $\calO(\rp, \bp)$ 
change this initial sign; the resulting first harmonic $\calO_1(r_\perp, \Delta_\perp)$ at smaller $x$
is shown above in Fig.~\ref{fig:odder}.\\

Finally, let us mention that a third $C$-conjugation odd amplitude corresponds to the exchange of a single
photon plus two gluons in a color singlet state. This corresponds to the expansion of
eq.~(\ref{eq:Sdipole-V-Vdagger}) to linear order in the electromagnetic and to quadratic order in
the color field. This contribution is expected to be small~\cite{Dumitru:2019qec} and has not been
considered here.

\section{BJKP--BLV Odderon}
\label{sec:linO}

\begin{figure}[htb]
  \begin{center}
  \includegraphics[scale = 0.7]{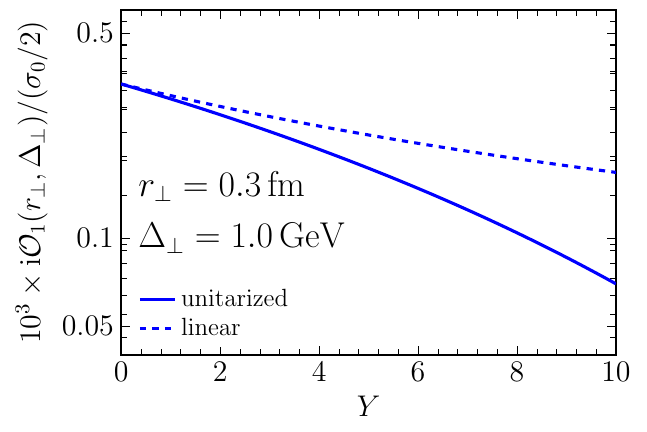}
  \includegraphics[scale = 0.7]{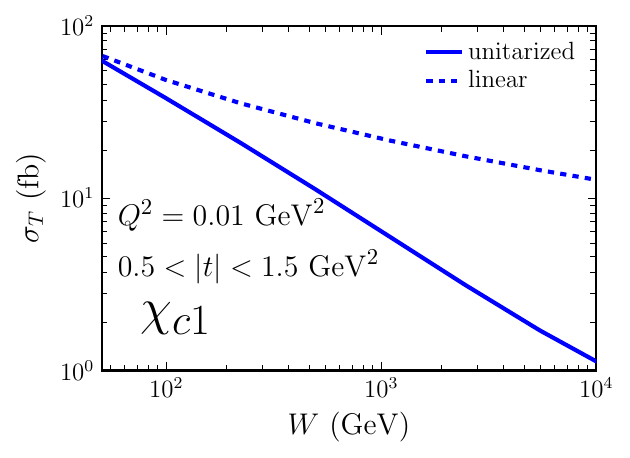}
  \end{center}
  \caption{Left: A comparison of the unitarized vs linear evolution of the Odderon amplitude $\calO_1(r_\perp, \Delta_\perp)$ as a function of $Y$. Right: A plot of the resulting $W$ dependence of the Odderon component of the $\gamma^* p \to \chi_{c1} p$ cross section.}
  \label{fig:bfklbk}
\end{figure}
In this work the small-$x$ evolution of the Odderon exchange amplitude has been computed using the non-linear evolution of the Odderon coupled to the BK equation \cite{Kovchegov:2003dm,Hatta:2005as} -- the ``unitarized solution''. It is interesting to compare to the BLV solution \cite{Bartels:1999yt} obtained from the BKP equation for the Odderon, that is, without including the unitarity corrections\footnote{Alternatively, one could solve for the BKP Green's function and reconstruct the Odderon amplitude from it~\cite{Chachamis:2016ejm}.} -- the ``linear solution".  Starting from the initial condition at $Y = 0$ obtained in \cite{Dumitru:2022ooz} and described in Sec.~\ref{sec:Odderon-Evol}, in Fig.~\ref{fig:bfklbk} we show $\calO_1(r_\perp,\Delta_\perp)$ as a function of $Y = \log(x_0/x_\calP)$ (left) with fixed $r_\perp = 0.3$ fm and $\Delta_\perp = 1$ GeV and up to $Y = 10$. The unitarized (full) and the linear (dashed line) results for $\calO_1(r_\perp,\Delta_\perp)$ both drop as a function of $Y$. For the unitarized solution this is more pronounced due to the non-linear corrections suppressing the Odderon exchange amplitude \cite{Motyka:2008ac,Lappi:2016gqe,Hatta:2017cte}. 

The impact of the unitarized vs linear evolution on the $W$-dependence of the Odderon component of the cross section is shown in Fig.~\ref{fig:bfklbk}(right). As an example, we consider the $\sigma_T(\gamma^* p \to \chi_{c1} p)$ cross section with the kinematic cuts as in Fig.~\ref{fig:csTW} from which we reproduced the result of unitarized evolution. 
The range in $Y$ shown in Fig.~\ref{fig:bfklbk}(left) roughly translates to the range in $W$ shown on the right. 
Even though linear evolution does lead to a much milder $W$-dependence of $\sigma_T$, the cross section does not seem to fully reach the BLV asymptotics (a constant value) even for $W = 10^4$ GeV. Note, however, that from the standard saddle point analysis one expects a $1/Y^3$ correction to the asymptotic behavior of the non-forward BLV exchange cross section -- as established for the analogous case of the non-forward BFKL cross section \cite{Forshaw:1995ax}. From the perspective of the EIC, taking the top $\sqrt{S} = 140$ GeV collision energy of the $ep$ system, $W$ can reach at most $W \approx 136$ GeV (at $y = 0.95$) resulting in a factor of $2-3$ difference in the cross section at the high~$y$ end of the photon flux. Due to the $1/y$ leading behavior of the photon flux, however, the exclusive $\chi_{cJ}$ production at the EIC will be dominated by much lower $W$, close to the experimental lower cutoff on $W$ for exclusive events, say for $2<Y<4$, where the unitarity corrections to the Odderon exchange cross section are below 30\% -- i.e.\ small in comparison to the theoretical uncertainty. 

\typeout{}
\bibliography{references}

\end{document}